\documentclass{aa}
\usepackage{graphicx}

\def\jh{\mbox{$\rm (J-H)$}}
\def\mMJ{\mbox{$\rm (m-M)_J$}}
\def\ebv{\mbox{$\rm E(B-V)$}}
\def\ejh{\mbox{$\rm E(J-H)$}}
\def\rc{\mbox{$\rm R_{core}$}}
\def\rl{\mbox{$\rm R_{lim}$}}
\def\rt{\mbox{$\rm R_{tidal}$}}
\def\ms{\mbox{$\rm M_\odot$}}
\def\ds{\mbox{$\rm d_\odot$}}
\def\dgc{\mbox{$\rm d_{GC}$}}
\def\jj{\mbox{$\rm J$}}
\def\hh{\mbox{$\rm H$}}
\def\ks{\mbox{$\rm K_S$}}

\def\mevol{\mbox{$\rm m_{evol}$}}
\def\mobs{\mbox{$\rm m_{obs}$}}

\def\mtot{\mbox{$\rm m_{tot}$}}
\def\mb{\mbox{$\rm m_{break}$}}
\def\kms{\mbox{$\rm km\,s^{-1}$}}
\def\tr{\mbox{$\rm t_{relax}$}}
\def\tcr{\mbox{$\rm t_{cross}$}}

\begin{document}

\title{Detailed analysis of open clusters: a mass function break and evidence of a fundamental 
plane}

\author{C. Bonatto\inst{1}  \and E. Bica\inst{1}}

\offprints{Ch. Bonatto - charles@if.ufrgs.br}

\institute{Universidade Federal do Rio Grande do Sul, Instituto de F\'\i sica, 
CP\,15051, Porto Alegre 91501-970, RS, Brazil\\
\mail{}
}

\date{Received --; accepted --}

\abstract{We derive photometric, structural and dynamical evolution-related parameters of 11 nearby 
open clusters with ages in the range 70\,Myr to 7\,Gyr and masses in the range $\approx400$\,\ms\ to 
$\approx5\,300$\,\ms. The clusters are homogeneously analysed by means of \jj, \hh\ and \ks\ 2MASS 
photometry, which provides spatial coverage wide enough to properly take into account the contamination 
of the cluster field by Galaxy stars. Structural parameters such as core and limiting radii are derived 
from the background-subtracted radial density profiles. Luminosity and mass functions (MFs) are built 
for stars later than the turnoff and brighter than the 2MASS PSC 99.9\% completeness limit. The total 
mass locked up in stars in the core and the whole cluster, as well as the corresponding mass densities, 
are calculated by taking into account the observed stars (evolved and main sequence) and extrapolating 
the MFs down to the H-burning mass limit, 0.08\,\ms. We illustrate the methods by analysing for the 
first time in the near-infrared the populous open clusters NGC\,2477 and NGC\,2516. For NGC\,2477 we 
derive an age of $1.1\pm0.1$\,Gyr, distance from the Sun $\ds=1.2\pm0.1$\,kpc, core radius $\rc=1.4\pm0.1$\,pc, 
limiting radius $\rl=11.6\pm0.7$\,pc and total mass $\mtot\approx(5.3\pm1.6)\times10^3$\,\ms. Large-scale 
mass segregation in NGC\,2477 is reflected in the significant variation of the MF slopes in different 
spatial regions of the cluster, and in the large number-density of giant stars in the core with respect 
to the cluster as a whole. For NGC\,2516 we derive an age of $160\pm10$\,Myr, $\ds=0.44\pm0.02$\,kpc, 
$\rc=0.6\pm0.1$\,pc, $\rl=6.2\pm0.2$\,pc and $\mtot\approx(1.3\pm0.2)\times10^3$\,\ms. Mass-segregation 
in NGC\,2516 shows up in the MFs. Six of the 11 clusters present a slope break in the MF occurring at 
essentially the same mass as that found for the field stars in Kroupa's universal IMF. The MF break is 
not associated to cluster mass, at least in the clusters in this paper. In two clusters the low-mass end of the MF 
occurs above the MF break. However, in three clusters the MF break does not occur, at least for
the mass range $\rm m\geq0.7\,\ms$. One possibility is 
dynamical evolution affecting the MF slope distribution. We also search for relations of structural and 
evolutionary parameters with age and Galactocentric distance. The main results for the present sample 
are: {\em (i)} cluster size correlates both with age and Galactocentric distance; {\em (ii)} because of 
size and mass scaling, core and limiting radii, and core and overall mass correlate; {\em (iii)} massive 
($\rm m\geq1\,000$\,\ms) and less-massive clusters follow separate correlation paths on the plane core 
radius and overall mass; {\em (iv)} MF slopes of massive clusters are restricted to a narrow range, while 
those of the less-massive ones distribute over a wider range. Core and overall MF flattening is related 
to the ratio ($\tau$) of age to relaxation time. For large values of $\tau$ the effects of large-scale 
mass segregation and low-mass stars evaporation can be observed in the MFs. In this sense, $\tau$ appears 
to characterize the evolutionary state of the clusters. We conclude that appreciable slope flattenings in 
the overall MFs of the less-massive clusters take $\sim6$ times longer to occur than in the core, while in 
the massive clusters they take a time $\sim13$ times longer. We investigate cluster parameters equivalent 
to those determining the fundamental plane of ellipticals. These parameters are: overall mass, projected 
mass density and core radius. We conclude that in the present sample there is evidence of a fundamental 
plane. Larger samples are necessary to pin down this issue.

\keywords{(Galaxy:) open clusters and associations: general} }

\titlerunning{Multi-parametric analysis of open clusters}

\authorrunning{C. Bonatto \and E. Bica}

\maketitle

\section{Introduction}
\label{intro}

Open clusters are self-gravitating stellar systems which span a broad range of ages and masses, 
and present significant variations in linear dimensions and morphology (e.g. Lyng\aa\ \cite{Lynga82}; 
Binney \& Merrifield \cite{Binney1998}; Tadross et al. \cite{Tad2002}). The structure of most open 
clusters can be described by two subsystems, the dense core and the sparse halo (corona). The stellar 
content of an open cluster presumably formed at the same time from the same parent interstellar 
cloud, thus sharing similar initial conditions. Stellar and dynamical evolution, as well as external 
interactions with Galactic structures, change the morphology and the internal mass distribution 
of clusters (Bergond, Leon \& Guibert \cite{Bergond2001}). As a consequence of large-scale mass 
segregation high-mass stars (as well as multiple systems) tend to be more concentrated in the core 
of evolved clusters while low-mass stars are transferred to the halo. In the oldest clusters (or the 
less-massive ones) the low-mass stars may evaporate from the cluster and get dispersed into the  
surroundings (de la Fuente Marcos \cite{delaF98}). Consequently, changes which occur along time 
must be reflected on observational parameters such as core and overall radii, the mass 
locked up in stars in the core and in the whole cluster, and the mass densities in both structures. In 
addition, core and overall mass-function slopes change as the clusters evolve (e.g. Kroupa 
\cite{Kroupa2001}). In this sense, open clusters are excellent laboratories in which models of star 
formation, stellar evolution and dynamical evolution can be tested both on theoretical grounds and  
numerical simulations. 

The relatively small number of stars (as compared to globular clusters) in open clusters poses some 
technical challenges to N-body simulations, particularly associated to the {\it (i)} increased 
granularity of the gravitational potential, {\it (ii)} significant fraction of binaries formed along 
the evolution, {\it (iii)} significant mass loss during the course of the normal stellar evolution, 
which affects cluster life-time, and {\it (iv)} non-spherical spatial shape. A discussion on the above 
points is in e.g. de la Fuente Marcos \& de la Fuente Marcos (\cite{delaF2002}). Consequently, it is 
important to search for correlations among observational parameters in order to better constrain 
theoretical studies and simulation codes. Accordingly, large spatial coverage, photometric uniformity 
and a homogeneous analysis method are essential to obtain a consistent set of intrinsic parameters. 
This point is important since most open clusters are projected against the Galactic plane and their 
fields may be heavily contaminated by Galactic (field) stars.

Previous searches for correlations involving open cluster parameters can be found in, e.g. Lyng\aa\ 
(\cite{Lynga82}), which is a compilation of literature data, Tadross et al. (\cite{Tad2002}), which 
is based on UBV CCD photometry, and Nilakshi, Pandey \& Mohan (\cite{Nilakshi2002}), which is based 
on star counts extracted from Digital Sky Survey plates. Because of the limited spatial coverage and/or 
limited photometrical depth of the previous observational techniques, most parameters were restricted 
to the central regions of the clusters and consequently, they suffered from poor background 
contamination subtraction, as well as cluster field crowding.

Parameters used in the present work are: {\it (i)} Galactocentric distance (\dgc), which is an 
external parameter; {\it (ii)} cluster age, which is associated to the stellar evolution; {\it (iii)} 
core (\rc) and limiting (\rl) radii, which are structural parameters associated to the dynamical 
evolution; {\it (iv)} core and overall mass; {\it (v)} core and overall projected mass density (in 
$\rm\ms\,pc^{-2}$); {\it (vi)} core and overall mass density (in $\rm\ms\,pc^{-3}$). Parameters 
listed in {\it (ii)} and {\it (iv)} to {\it (vi)} are intrinsically associated to the stellar and 
dynamical evolution. We refer to the whole structure ($\rm 0\leq r\leq\rl$) of a cluster as overall.

For the sake of spatial and photometric uniformity, we employ in the present work \jj, \hh\ and \ks\ 
2MASS\footnote {The Two Micron All Sky Survey, All Sky data release (Skrutskie et al. \cite{2mass1997}), 
available at {\em http://www.ipac.caltech.edu/2mass/releases/allsky/}} photometry. The 2MASS Point 
Source Catalogue (PSC) is uniform reaching relatively faint magnitudes covering nearly all the sky, 
allowing a proper background definition even for clusters with large angular sizes. The clusters in the 
present work have already been studied by us in previous papers (in proper motion and dynamical state 
studies), except for the rich open clusters NGC\,2477 and NGC\,2516, which are studied in the 
near-infrared for the first time and illustrate in detail the present analysis method as well. 

Age and Galactocentric distance (derived from the distance from the Sun) are obtained by fitting 
isochrones to the near-infrared 2MASS colour-magnitude diagrams (CMDs), which gives 
as well the foreground colour excess in that direction (e.g. Bica, Bonatto \& Dutra \cite{BBD2004a},
\cite{BBD2004b}). The statistical significance of these parameters depends directly on the 
quality and depth of the photometry (Bonatto, Bica \& Pavani \cite{BBP2004}; Bonatto, Bica \& Santos 
Jr. \cite{BBS2005}). The core radius is derived by fitting a King model to the radial density profile, 
while the limiting radius corresponds to the distance from the center where the radial density profile 
merges with the background level. In this case, deep enough photometry is necessary to produce a radial 
density profile which includes low main-sequence (MS) stars as well, while spatial coverage is 
fundamental, since a high background may produce an 
underestimated cluster dimension (Bica, Bonatto \& Dutra \cite{BBD2004b}; Bonatto, Bica \& Pavani 
\cite{BBP2004}; Bonatto, Bica \& Santos Jr. \cite{BBS2005}). Wide spatial coverage is  
important for the dynamically evolved clusters, since a significant fraction of the low-mass content 
may already have been dispersed into the surroundings (Bonatto \& Bica \cite{BB2003}; Bonatto, 
Bica \& Pavani \cite{BBP2004}; Bonatto, Bica \& Santos Jr. \cite{BBS2005}). Core and overall masses are 
obtained by first deriving the intrinsic luminosity (LF) and mass (MF) functions, which depend 
essentially on properly taking into account the stellar background distribution, and then integrating 
the MFs down to the H-burning limit stars. Consequently, depth and coverage are 
fundamental to properly derive mass, particularly for clusters projected against the disk and/or bulge 
(Bica, Bonatto \& Dutra \cite{BBD2004b}; Bonatto, Bica \& Santos Jr. \cite{BBS2005}). 

In the present work we intend to search for fundamental relations  in the parameter space described 
above. Additionally, since star clusters are, to a first approximation, scaled-down versions of 
elliptical galaxies, we intend to explore the possibility of an analogous fundamental plane (FP, e.g. 
Dressler et al. \cite{Dress1987}; Djorgovski \& Davis \cite{DD1987}) of open cluster parameters. The 
parameters which could correspond to those of the FP in ellipticals could be cluster overall and/or 
core mass, core radius (the analogous of the effective radius in the ellipticals), and the projected 
mass density. However, the conceptual differences between ellipticals and open clusters, e.g. total 
population and dynamical evolutionary state, may hinder such eventual relations.

This paper is organized as follows. In Sect.~\ref{ttc} we provide general data on the target clusters. 
In Sect.~\ref{PopOC} we illustrate the analysis methods with NGC\,2477 and NGC\,2516. Cluster parameters 
are presented and correlations are discussed in Sect.~\ref{MPA}. The MF slope break and dynamical effects 
on MFs are discussed in Sect.~\ref{MFT}. In Sect.~\ref{FP} we discuss a possible FP in open clusters. 
Finally, concluding remarks are given in Sect.~\ref{Conclu}. 

\section{The target clusters}
\label{ttc}

\begin{table*}
\caption[]{General data on the target clusters.}
\label{tab1}
\renewcommand{\tabcolsep}{0.35mm}
\begin{tabular}{lcccccccccccc}
\hline\hline
&\multicolumn{5}{c}{WEBDA}&\multicolumn{6}{c}{Present work}\\
\cline{2-6}\cline{8-13}\\
Cluster&$\alpha(2000)$&$\delta(2000)$&Age&\ebv&\ds&&$\alpha(2000)$&$\delta(2000)$&Age&\ebv&\ds&\dgc\\
&(hms)&($^\circ\arcmin\arcsec$)&(Myr)&&(kpc)&&(hms)&($^\circ\arcmin\arcsec$)&(Myr)&&(kpc)&(kpc)\\
(1)&(2)&(3)&(4)&(5)&(6)&&(7)&(8)&(9)&(10)&(11)&(12)\\
\hline
M\,26&$18:45:18$&$-09:23:00$    &85&0.59&1.60&&$18:45:14^\dag$&$-09:22:55^\dag$&$70\pm10$&$0.42\pm0.03$&$1.6\pm0.1$&6.6\\
NGC\,2516&$07:58:04$&$-60:45:12$&113&0.10&0.41&&$07:58:04$&$-60:45:12$&$160\pm10$&$0.00$&$0.44\pm0.02$&8.0\\
NGC\,2287&$06:46:01$&$-20:45:24$&243&0.03&0.69&&$06:46:01$&$-20:45:24$&$160\pm10$&$0.00$&$0.8\pm0.1$&8.5\\
M\,48&$08:13:43$&$-05:45:00$    &360&0.03&0.77&&$08:13:45^\dag$&$-05:46:42^\dag$&$360\pm40$&$0.00$&$0.8\pm0.1$&8.5\\
M\,93&$07:44:30$&$-23:51:24$    &387&0.05&1.04&&$07:44:31^\dag$&$-23:50:42^\dag$&$400\pm50$&$0.03\pm0.01$&$1.1\pm0.1$&8.6\\
NGC\,5822&$15:04:21$&$-54:23:48$&662&0.15&0.92&&$15:04:21$&$-54:23:48$&$1000\pm100$&$0.00$&$0.7\pm0.1$&7.4\\
NGC\,2477&$07:52:10$&$-38:31:48$&705&0.28&1.22&&$07:52:18^\dag$&$-38:31:57^\dag$&$1100\pm100$&$0.06\pm0.03$&$1.2\pm0.1$&8.4\\
NGC\,3680&$11:25:38$&$-43:14:36$&1200&0.07&0.94&&$11:25:38$&$-43:14:36$&$1600\pm100$&$0.00$&$1.0\pm0.1$&7.8\\
IC\,4651&$17:24:49$&$-49:56:00$ &1140&0.12&0.89&&$17:24:52^\dag$&$-49:56:52^\dag$&$1800\pm200$&$0.00$&$0.9\pm0.1$&7.2\\
M\,67 &$08:51:18$&$+11:48:00$   &2560&0.06&0.91&&$08:51:16^\dag$&$+11:48:54^\dag$&$3200\pm100$&$0.00$&$0.9\pm0.1$&8.7\\
NGC\,188&$00:47:28^\dag$&$+85:15:18^\dag$&4300 &0.08&2.05 &&$00:47:53$&$+85:15:30$&$7000\pm1000$&$0.00$&$1.7\pm0.1$&8.9\\
\hline
\end{tabular}
\begin{list}{Table Notes.}
\item Colour excesses in col.~10 are derived from the isochrone fit to the 2MASS CMDs. 
Uncertainty in Galactocentric distance in col.~12 is $\approx0.1$\,kpc for all clusters.
Uncertainties in cols.~9 to 12 are derived from the isochrone fit. ($^\dag$) Coordinates 
optimized to maximize the number-density of stars in the cluster center (Sect.~\ref{PopOC}).
\end{list}
\end{table*}

We selected for the present work 9 open clusters already studied 
by us with 2MASS in previous papers for different goals: M\,26 (NGC\,6694), NGC\,2287, M\,48 
(NGC\,2548), M\,93 (NGC\,2447), NGC\,5822, NGC\,3680, IC\,4651, M\,67 (NGC\,2682) and 
NGC\,188. Proper motions and binarity in M\,26, NGC\,2287, M\,48, M\,93, NGC\,5822, NGC\,3680, 
IC\,4651 and M\,67 were investigated in Bica \& Bonatto (\cite{BiBo2004}). The advanced 
dynamical state of NGC\,3680 was analysed in Bonatto, Bica \& Pavani (\cite{BBP2004}). Mass 
segregation in M\,67 was studied in Bonatto \& Bica (\cite{BB2003}). The dynamical state of 
the old open cluster NGC\,188 was investigated in Bonatto, Bica \& Santos Jr. (\cite{BBS2005}).

The populous open clusters NGC\,2477 and NGC\,2516 are discussed in Sect.~\ref{PopOC}. The clusters 
above span a wide age range, which is important since age is a fundamental factor associated to 
dynamical evolution. General data for the above open clusters are given in Table~\ref{tab1}. The data 
in cols.~2-6 are from the WEBDA\footnote{\em http://obswww.unige.ch/webda} open cluster 
database (Mermilliod \cite{Merm1996}). In cols.~7-12 we list parameters derived from the 2MASS 
photometry.

We indicate below papers and results relevant to the present work, discriminated by open
cluster. 

\paragraph{\tt M\,26:} Battinelli, Brandimarti \& Capuzzo-Dolcetta (\cite{BBCD94}) give a
limiting radius of $\rl\approx3.3$\,pc and a total stellar mass of $\mtot\approx240\,\ms$.

\paragraph{\tt NGC\,2516:} Battinelli, Brandimarti \& Capuzzo-Dolcetta (\cite{BBCD94}) give 
$\rl\approx1.9$\,pc and $\mtot\approx282\,\ms$. Bergond, Leon \& Guibert (\cite{Bergond2001}) 
give $\ds\approx0.4$\,kpc, an age of $\approx113$\,Myr, $\rt=1.9$\,pc and $\mtot\approx170$\,\ms. 
Tadross et al. (\cite{Tad2002}) give $\ds\approx0.4$\,kpc, $\rm age\approx63$\,Myr, 
$\rl\approx1.8$\,pc, $\dgc\approx8.5$\,kpc and $\mtot\approx36\,\ms$. Jeffries, Thurston \& Hambly 
(\cite{Jeff2001}) estimate an age of $\approx150$\,Myr and total mass of 
$\mtot\approx1240-1560$\,\ms. They also detected a sharp change in the MF slope for stars with 
$0.3\leq m(\ms)\leq0.7$.

\paragraph{\tt NGC\,2287:} Battinelli, Brandimarti \& Capuzzo-Dolcetta (\cite{BBCD94}) give 
$\rl\approx4.4$\,pc and $\mtot\approx195\,\ms$. Bergond, Leon \& Guibert (\cite{Bergond2001}) 
included this object in a study of the gravitational tidal effects on open clusters. They give 
a distance from the Sun of $\ds\approx0.7$\,kpc, an age of $\approx240$\,Myr, a tidal radius 
of $\rt=4.1$\,pc and a total stellar mass of $\mtot\approx300$\,\ms. 

\paragraph{\tt M\,48:} Bergond, Leon \& Guibert (\cite{Bergond2001}) give $\ds\approx0.8$\,kpc, 
an age of $\approx360$\,Myr and $\rt=4.8$\,pc.

\paragraph{\tt M\,93:} Nilakshi, Pandey \& Mohan (\cite{Nilakshi2002}) derived
$\rc\approx0.5$\,pc and $\rl\approx3.3$\,pc.

\paragraph{\tt NGC\,5822:} Battinelli, Brandimarti \& Capuzzo-Dolcetta (\cite{BBCD94}) give 
$\mtot\approx251\,\ms$.

\paragraph{\tt NGC\,2477:} Based on UBV photometry of 83 red-giant candidates in the field of 
NGC\,2477, Eigenbrod et al. (\cite{Eigenbrod2004}) derived a core radius $\rc=1.8$\,pc and a 
tidal radius $\rt=8.1$\,pc. They also estimate that the total mass locked up in stars more 
massive than $\approx1$\,\ms\ must be larger than $5\,400$\,\ms. A break in the mass function 
in the lower main sequence at $\rm m\approx0.27$\,\ms\ was suggested by von Hippel et al. 
(\cite{Hippel1996}). Nilakshi, Pandey \& Mohan (\cite{Nilakshi2002}) derived $\rc\approx2.2$\,pc 
and $\rl\approx9.1$\,pc. Tadross et al. (\cite{Tad2002}) give $\ds\approx1.4$\,kpc, 
$\rm age\approx1$\,Gyr, $\rl\approx5.5$\,pc, $\dgc\approx9.0$\,kpc and $\mtot\approx2100\,\ms$.

\paragraph{\tt NGC\,3680:} Bonatto, Bica \& Pavani (\cite{BBP2004}) studied this dynamically 
advanced open cluster and derived $\ds\approx1$\,kpc, age $\approx1.6$\,Gyr, $\rc\approx0.7$\,pc,
$\rl\approx6.4$\,pc and $\mtot\approx550\,\ms$.

\paragraph{\tt IC\,4651:} Meibom, Andersen \& Nordstr\"om (\cite{Meibom2002}) derived an age of 
$\approx1.7$\,Gyr, $\rt\approx5.8$\,pc and a present total mass of $\sim630\,\ms$; they also 
present evidence of moderate mass segregation. Tadross et al. (\cite{Tad2002}) give $\ds\approx0.8$\,kpc, 
$\rm age\approx2.2$\,Gyr, $\rl\approx1.4$\,pc, $\dgc\approx7.8$\,kpc and $\mtot\approx100\,\ms$.
 
\paragraph{\tt M\,67:} Battinelli, Brandimarti \& Capuzzo-Dolcetta (\cite{BBCD94}) give 
$\mtot\approx407\,\ms$. Bonatto \& Bica (\cite{BB2003}) derived an age of $\approx3.2$\,Gyr,
$\ds\approx0.9$\,kpc, $\rc=1.2$\,pc and $\rl\approx1.8$\,pc. They also detected mass segregation.
Mass segregation in M\,67 was also detected by Sarajedini et al. (\cite{Sarajedini1999}).
Montgomery, Marschall \& Janes (\cite{Montg1993}) estimate a mass of $\approx800\,\ms$.
Detailed N-body modelling of the dynamical and stellar evolution of M\,67 was carried out 
by Hurley et al. (\cite{Hurley2001}); they estimate a much larger initial mass for M\,67.
Tadross et al. (\cite{Tad2002}) give $\ds\approx0.8$\,kpc, $\rm age\approx4$\,Gyr, 
$\rl\approx3.2$\,pc, $\dgc\approx9.1$\,kpc and $\mtot\approx820\,\ms$.

\paragraph{\tt NGC\,188:} Bonatto, Bica \& Santos Jr. (\cite{BBS2005}) derived an age of 
$\approx7$\,Gyr, $\ds\approx1.7$\,kpc, $\rc\approx1.3$\,pc, $\rt\approx21$\,pc and a total 
mass of $\approx4\,000\,\ms$; they detected strong mass segregation. Mass segregation was also 
detected by Sarajedini et al. (\cite{Sarajedini1999}). Nilakshi, Pandey \& Mohan (\cite{Nilakshi2002}) 
derived $\rc\approx2.1$\,pc and $\rl\approx6.5$\,pc. Tadross et al. (\cite{Tad2002}) give $\ds=1.5$\,kpc, 
$\rm age\approx5$\,Gyr, $\rl\approx3.1$\,pc, $\dgc\approx9.4$\,kpc and an exceedingly low
mass of $\approx208\,\ms$.

The near-infrared CMDs and radial density profiles of M\,26, NGC\,2287, M\,48, M\,93, NGC\,5822, 
NGC\,3680, IC\,4651 and M\,67 are shown in Bica \& Bonatto (\cite{BiBo2004}). For NGC\,188 they 
are in Bonatto, Bica \& Santos Jr. (\cite{BBS2005}). 

Comparing WEBDA compilation with the present results (Table~\ref{tab1}), ages and distances from
the Sun agree in general. However, the colour excesses \ebv\ tend to be smaller when derived
from the near-infrared, as discussed in Bica \& Bonatto (\cite{BiBo2004}).

\section{The populous open clusters NGC\,2477 and NGC\,2516}
\label{PopOC}

\subsection{The intermediate-age open cluster NGC\,2477}
\label{n2477}

According to WEBDA, the central coordinates of NGC\,2477 are (J2000) $\alpha=07^h52^m10^s$, and 
$\delta=-38^\circ31\arcmin48\arcsec$. However, the corresponding radial density profile (built with a 
step in radius of $\rm\Delta r=1\arcmin$) presented a dip for $\rm r=0\arcmin$. We searched for a new 
center by examining histograms for the number of stars in bins of right ascension and declination. The 
resulting coordinates which maximize the density of stars at the center are (J2000) $\alpha=07^h52^m18^s$, 
and $\delta=-38^\circ31\arcmin57\arcsec$, corresponding to $\ell=253.58^\circ$ and $b=-5.82^\circ$. 
In what follows we refer to these coordinates as the center of NGC\,2477.

\subsubsection{The 2MASS photometry and near-infrared CMD}
\label{2mass}

The VizieR\footnote{\em http://vizier.u-strasbg.fr/viz-bin/VizieR?-source=II/246} tool has 
been used to extract 2MASS photometry of the stars in a circular area with radius $\rm r=60\arcmin$ 
centered on the optimized coordinates (Table~\ref{tab1}). 2MASS photometric uncertainties
as a function of magnitude are discussed in Soares \& Bica (\cite{SB2002}) and Bonatto, Bica,
\& Santos Jr. (\cite{BBS2005}).

\begin{figure} 
\resizebox{\hsize}{!}{\includegraphics{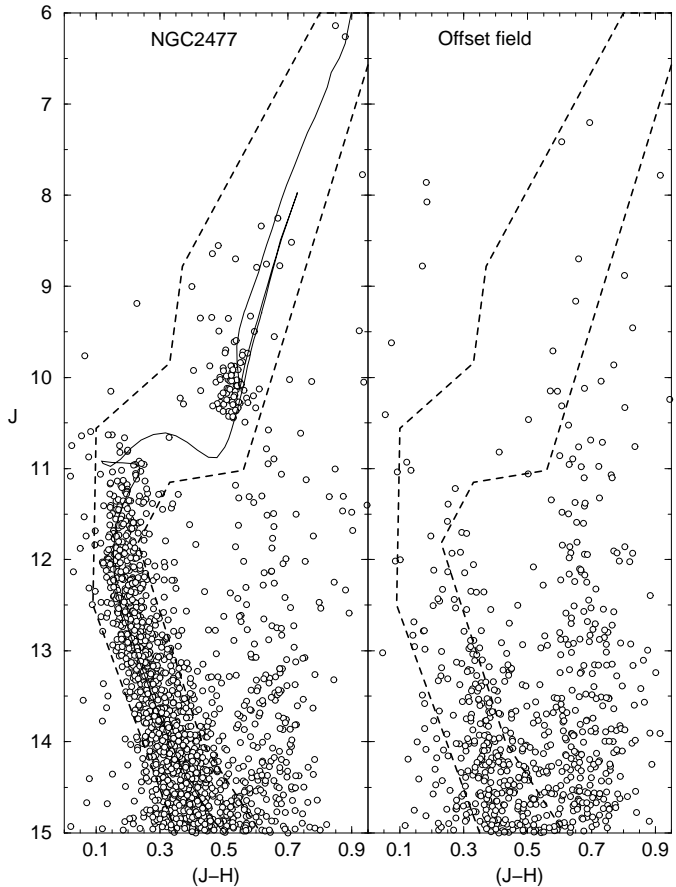}}
\caption[]{Left panel: $\jj\times\jh$ CMD of the central 10\arcmin\ ($\sim3.6$\,pc) of NGC\,2477
with the {\em best-fit} Padova isochrone ($\rm age\sim1.1$\,Gyr) superimposed. Note the densely
populated main sequence and rich giant clump. Right panel: CMD of the corresponding (same area) 
offset field. The colour-magnitude filter is shown in both panels. }
\label{fig1}
\end{figure}

Because of the  moderately high Galactic latitude ($b\approx-6^\circ$), the field of NGC\,2477 
presents limited background contamination. This can be seen in Fig.~\ref{fig1} in which we show the 
$\jj\times\jh$ CMD of the central ($\rm r=10\arcmin\sim3.6$\,pc) region of NGC\,2477 (left panel) 
compared with the corresponding, same area, offset field (right panel). To match the areas
of both CMDs in Fig.~\ref{fig1} the offset field has been obtained by extracting stars in the 
outermost ring, located at $\rm 59.16\arcmin\leq r\leq 60\arcmin$, which represents essentially the 
Galactic background/foreground stellar contribution. The CMD of the central region of NGC\,2477 
presents a well-defined morphology including a densely populated MS up to the turnoff and 
a rich giant clump that strongly contrasts with the CMD of the offset field. We show in both 
panels the colour-magnitude filter which is used to increase membership probability by discarding 
stars with far-off colours with respect to the isochrone fit (see below).

Cluster parameters have been derived by fitting solar metallicity Padova isochrones (Girardi 
et al. \cite{Girardi2002}) computed with the 2MASS \jj, \hh\ and \ks\ filters\footnote{\em\tiny 
http://pleiadi.pd.astro.it/isoc\_photsys.01/isoc\_photsys.01.html},
to the observed CMD in Fig.~\ref{fig1}. The 2MASS transmission filters produced isochrones 
very similar to the Johnson ones, with differences of at most 0.01 in \jh\ (Bonatto, Bica \& 
Girardi \cite{BBG2004}). For reddening and absorption transformations we use R$_V$ = 3.2, and 
the relations A$_J = 0.276\times$A$_V$ and $\ejh=0.33\times\ebv$, according to Dutra, Santiago 
\& Bica (\cite{DSB2002}) and references therein. The {\em best-fit} for NGC\,2477 uses the
$1.12$\,Gyr isochrone. Taking into account the uncertainties associated to
the isochrone fit we derive for NGC\,2477 an age of $1.1\pm0.1$\,Gyr, a colour excess
$\ejh=0.02\pm0.01$ which corresponds to $\ebv=0.06\pm0.03$, a distance modulus $\mMJ=10.5\pm0.1$,
and a distance from the Sun $\ds=1.23\pm0.06$\,kpc. With this value the Galactocentric distance
of NGC\,2477 turns out to be $\dgc=8.4\pm0.1$\,kpc, using 8.0\,kpc as the distance of the Sun
to the center of the Galaxy (Reid \cite{Reid93}).

\subsubsection{Structural parameters}
\label{struc}

Structural parameters of NGC\,2477 have been derived by means of the radial density profile, defined 
as the projected number of stars per area around the cluster center. Before counting stars we applied 
the colour filter shown in Fig.~\ref{fig1} to the CMD of the cluster (for stars with distance to the 
center from 0\arcmin\ to 60\arcmin) in order to discard most of the contamination by Milky Way stars. 
This procedure has been previously applied in the analysis of the open clusters M\,67 (Bonatto \& Bica 
\cite{BB2003}), NGC\,188 (Bonatto, Bica \& Santos Jr. \cite{BBS2005}) and NGC\,3680 (Bonatto, Bica \& 
Pavani \cite{BBP2004}). To further minimize the probability of background contamination, spurious 
detections and the increase of photometric uncertainties at faint magnitudes we restricted the radial 
density analysis to filtered stars with $\jj\leq14.5$, corresponding to $\rm M_J\leq4.0$ and mass 
$\rm m\geq0.95\,\ms$ (Sect.~\ref{MF}). At $\jj=14.5$ the photometric uncertainties amount to 
$\rm\epsilon_J\approx0.04$ (Bonatto, Bica \& Santos Jr. \cite{BBS2005}). The radial density 
profile was obtained by counting stars inside concentric annuli with a step of 1.0\arcmin\ in 
radius. The background contribution level corresponds to the average number of stars included in 
the ring located at $\rm 40\arcmin\leq r\leq 60\arcmin$ ($\rm14.3\leq r(pc)\leq21.5$). The above 
procedures were applied to the filtered CMD.

\begin{figure} 
\resizebox{\hsize}{!}{\includegraphics{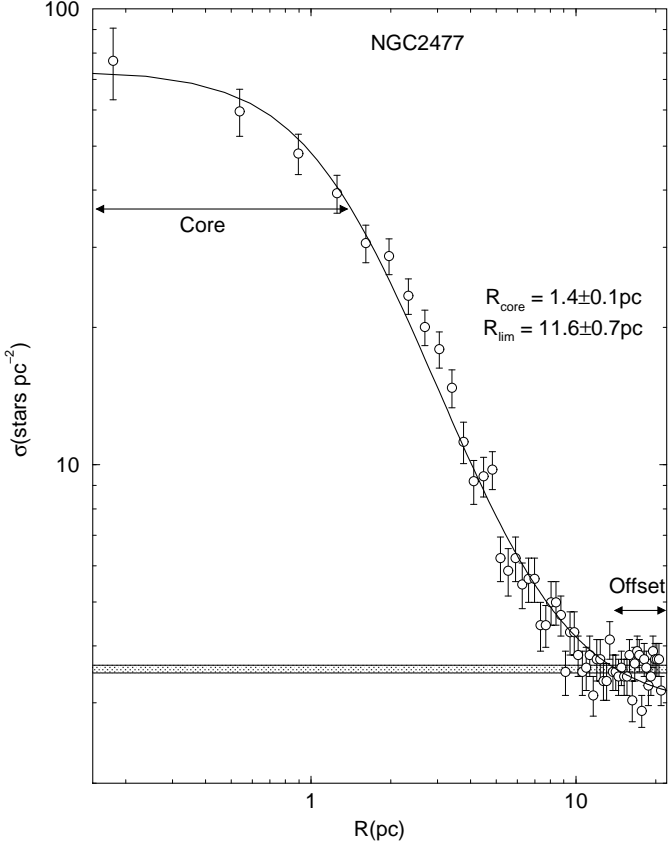}}
\caption[]{Projected radial density profile of stars in NGC\,2477. The average background level is 
shown as the narrow shaded rectangle; $1\sigma$ Poisson errors are also shown. The solid line shows 
the two-parameter King model fit to the radial profile. The extent of the core region and that of the 
offset field are indicated.}
\label{fig2}
\end{figure}

The resulting radial density profile of NGC\,2477 is shown in Fig.~\ref{fig2}. For 
absolute comparison between clusters we scale the radius in the abscissa in parsecs, and the 
number density of stars in the ordinate in $\rm stars\,pc^{-2}$\ using the distance derived 
in Sect.~\ref{2mass}. The statistical significance of the profile is reflected in the relatively 
small $1\sigma$\ Poisson error bars.

Structural parameters of NGC\,2477 were derived by fitting the two-parameter King (\cite{King1966a}) 
surface density profile to the background-subtracted radial distribution of stars. The two-parameter 
King model essentially describes the intermediate and central regions of normal clusters (King 
\cite{King1966b}; Trager, King \& Djorgovski \cite{TKD95}). The fit was performed using a nonlinear 
least-squares fit routine which uses the errors as weights. The best-fit solution is shown in 
Fig.~\ref{fig2} superimposed on the radial density profile. From the fit we derived a core radius 
$\rc=1.4\pm0.1$\,pc and a projected central density of stars of $\rm\sigma_0\approx70\pm7\,stars\,pc^{-2}$. 
Our value of \rc\ is somewhat smaller than $\rc=1.8$\,pc of Eigenbrod et al. (\cite{Eigenbrod2004}) 
and $\rc=2.2$\,pc of Nilakshi, Pandey \& Mohan (\cite{Nilakshi2002}). The present central density of
stars is $\sim3\times$ that of Nilakshi, Pandey \& Mohan (\cite{Nilakshi2002}).

Within uncertainties the two-parameter King model reproduces well the radial density profile in 
NGC\,2477. Since it follows from an isothermal (virialized) sphere, the close similarity 
of the radial distribution of stars in NGC\,2477 with a King profile suggests that despite its 
intermediate age, this cluster has evolved into an advanced dynamical state. We will return 
to this point in Sect.~\ref{MF}. 

Considering the radial density profile fluctuations with respect to the background level, we can 
define a limiting radius (\rl) for the cluster, in the sense that for regions beyond $\rl$, the 
null-contrast between cluster and background star density would produce large Poisson 
errors and consequently, meaningless results. Thus, for practical purposes, the 
bulk of the cluster's stars are contained within $\rl$. In the case of NGC\,2477 the limiting radius 
turns out to be $\rl=11.6\pm0.7$\,pc. Finally, we derived the tidal radius (\rt) by 
fitting the three-parameter King (\cite{King1962}) model to the background-subtracted radial 
distribution of stars in Fig.~\ref{fig2}. The resulting $\rt=16\pm2$\,pc is $\sim33\%$\ larger than 
\rl. Our value of \rt\ is $\sim2$\ times larger than the values given by Eigenbrod et al. 
(\cite{Eigenbrod2004}) and Nilakshi, Pandey \& Mohan (\cite{Nilakshi2002}), and a factor of $\sim3$\ 
larger than that of Tadross et al. (\cite{Tad2002}). The significant differences between our values of 
\rc\ and \rt\ and those of previous works may be accounted for by the depth of the 2MASS photometry,
the proper background definition and the use of a colour-magnitude filter.

\subsubsection{Mass functions}
\label{MF}

The populous nature of NGC\,2477 provides an opportunity to study in detail with a high statistical
significance the spatial distribution of mass functions $\left(\phi(m)=\frac{dN}{dm}\right)$. Based 
on the King profile fit and taking into account the features present in the radial 
density profile of NGC\,2477 (Fig.~\ref{fig2}) we selected the following regions: {\em (i)} $\rm 
0.0\leq r(pc)\leq1.4$ (core), {\em (ii)} $\rm 1.4\leq r(pc)\leq5.2$, {\em (iii)} $\rm 5.2\leq 
r(pc)\leq11.6$\ and {\em (iv)} $\rm 0.0\leq r(pc)\leq11.6$ (overall). Regions {\em (ii)} and {\em (iii)} 
are the halo of NGC\,2477. In order to maximize the significance of 
background counts, we consider as offset field the outermost ring at $\rm 14.3\,pc\leq r\leq 
21.5\,pc$, which lies $\sim2.7$\,pc beyond the cluster's limiting radius. 

\begin{figure} 
\resizebox{\hsize}{!}{\includegraphics{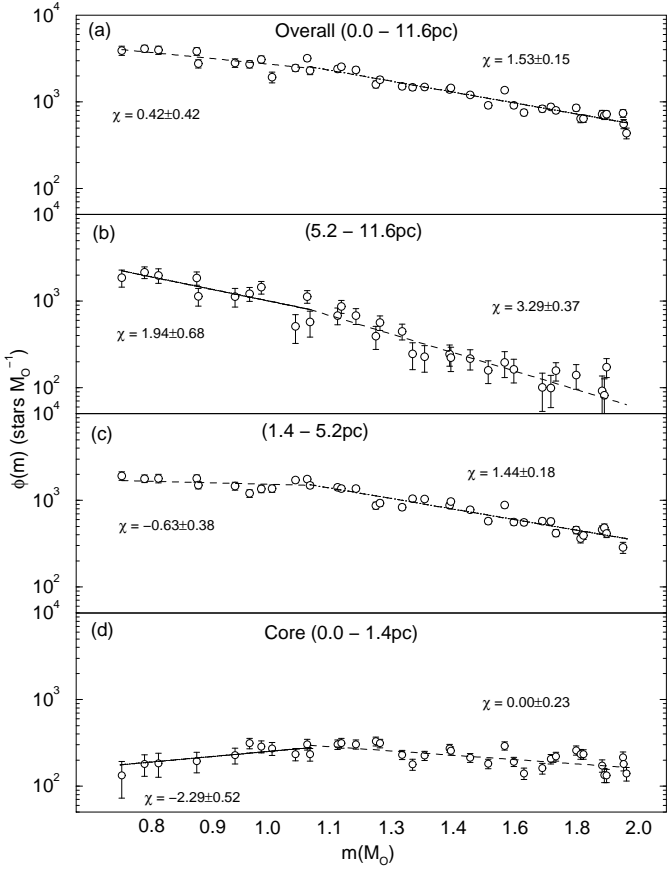}}
\caption[]{Mass functions in different spatial regions of NGC\,2477. Each panel contains MFs derived 
from the \jj, \hh\ and \ks\ 2MASS photometry. A break in the MFs at $\rm m\approx1.08$\,\ms\ can be
seen particularly in panels (c) and (d). MF fits $\left(\phi(m)\propto m^{-(1+\chi)}\right)$ for each
mass range are shown as dashed lines. The respective MF slopes are given. Note that the flattest 
(panel (d)) and steepest (panel (b)) MFs coincide spatially in both mass ranges.}
\label{fig3}
\end{figure}

Although relatively small, the Galactic stars contamination of the CMD of NGC\,2477 must be taken 
into account in order to derive the intrinsic luminosity and mass distributions of the member stars. 
To do this we first apply the colour filter (Fig.~\ref{fig1}) to both cluster and offset field CMDs. 
The filtering process takes into account most of the background, leaving a residual contamination. 
We deal with this residual contamination statistically by building the LFs for each cluster region 
and offset field. We build LFs for the three 2MASS bands independently, taking into account the 99.9\% 
Point Source Catalogue Completeness Limit\footnote{Corresponding to the Level\,1 Requirement, according 
to {\em\tiny http://www.ipac.caltech.edu/2mass/releases/allsky/doc/sec6\_5a1.html }}. Consequently, 
the faint magnitude limit of each LF is $\jj=15.8$, $\hh=15.1$\ and $\ks=14.3$, respectively. We take 
the turnoff as the bright limit to avoid inconsistencies in the mass-luminosity relation. For each 
2MASS band we build a LF by counting stars in magnitude bins from the respective faint magnitude limit 
to the turnoff, both for each cluster region and offset field. Considering that the solid angle of the 
offset field is different from that of a given cluster region, we multiply the offset field LF by a 
numerical factor so that the solid angles match. The intrinsic LF of each cluster region is obtained 
by subtracting the respective (i.e. solid angle-corrected) offset field LF from that of the cluster
region. Finally, the intrinsic LFs are transformed into MFs using the mass-luminosity relation obtained 
from the 1.12\,Gyr Padova isochrone and the distance modulus $\mMJ=10.5$. Remark that these procedures 
are repeated independently for the three 2MASS bands. The final MF of a given cluster region is produced 
by combining the \jj, \hh\ and \ks\ MFs into a single MF. The resulting spatial MFs of NGC\,2477, 
covering the mass range $\rm0.76\leq m(\ms)\leq1.95$, are shown in Fig.~\ref{fig3}.

The MFs present a break followed by slope flattening for masses in the range $\rm0.76\leq 
m(\ms)\leq1.08$, which is noticeable particularly in the core (panel (d)) and in the region 
$\rm 1.4\leq r(pc)\leq 5.2$\ (panel (c)). We note that von Hippel et al. (\cite{Hippel1996}), 
based on deep V and I HST and WFPC2 photometry of the central parts of NGC\,2477, described a 
break in the MF, but occurring at a lower mass value, $\rm m\approx0.27$\,\ms\ (Sect.~\ref{MFT}). 

Bearing in mind the MF break at $\mb=1.08$\,\ms\ we fit the function $\phi(m)\propto m^{-(1+\chi)}$\ in 
the mass ranges $\rm 0.76\leq m(\ms)\leq 1.08$\ and $\rm 1.08\leq m(\ms)\leq 1.95$. The resulting
fits are shown in Fig.~\ref{fig3}, and the MF slopes of each mass range are given in cols.~4 and 5
of Table~\ref{tab2}, respectively.

We provide in Table~\ref{tab2} parameters derived from the LFs and MFs. The number of evolved stars 
(col.~2) is calculated by integrating the intrinsic LFs for magnitudes brighter than the turnoff. 
Multiplying this by the mass at the turnoff (1.98\,\ms) gives an estimate of the evolved-star mass 
(col.~3). This procedure produces a realistic value of the number of member evolved stars because
the background contamination has already been statistically subtracted from the LF. The MF slopes in 
the two mass ranges are in cols.~4 and 5. The number of MS stars and 
corresponding stellar mass are derived by integrating the MF from the faint magnitude limit to the 
turnoff. We add to these the corresponding values of the number and mass of evolved stars to derive 
the total number of observed stars (col.~6), observed mass (col.~7), projected mass density (col.~8) 
and mass density (col.~9). 

An estimate of the total mass locked up in stars in NGC\,2477 was made by taking into account all stars 
from the turnoff down to the H-burning mass limit, $0.08\,\ms$. We do this by directly extrapolating 
the low-mass MFs ($\rm 0.76\leq m(\ms)\leq 1.08$) down to $0.08\,\ms$, except for the region $\rm 
5.2\leq r(pc)\leq11.6$ for which the slope is $\chi=1.9\pm0.7$. In this case we assume the universal 
Initial Mass Function (IMF) of Kroupa (\cite{Kroupa2001}), in which $\chi=0.3\pm0.5$ for the range
$\rm 0.08\leq m(\ms)\leq0.5$. In the range $\rm 0.5\leq m(\ms)\leq0.76$ we use our value of $\chi$ 
which, within the uncertainty is similar to the value presented in Kroupa (\cite{Kroupa2001}) for the 
equivalent mass range. The resulting extrapolated values of the number of stars and extrapolated mass 
(added to the corresponding values of the number and mass of evolved stars) are given respectively in 
cols.~10 and 11 of Table~\ref{tab2}. The extrapolated projected and mass densities are in cols.~12 and 
13. 

The total mass locked up in stars in the core of NGC\,2477 results in 
$\rm\mtot(core)\approx(5.2\pm0.1)\times10^2$\,\ms, nearly $25\%$ larger than the mass corresponding
to the observed stars. The total overall mass is $\rm\mtot(overall)\approx(5.3\pm1.5)\times10^3$\,\ms, 
which is in close agreement with the mass estimate (for stars with $\rm m\geq1\,\ms$) of Eigenbrod 
et al. (\cite{Eigenbrod2004}) and $\sim2.5$\ times larger than the estimate of Tadross et al. 
(\cite{Tad2002}). The extrapolated overall mass is nearly twice the mass stored in the observed stars.

\begin{table*}
\caption[]{Data derived from the mass functions in NGC\,2477 and NGC\,2516.}
\begin{scriptsize}
\label{tab2}
\renewcommand{\tabcolsep}{0.57mm}
\renewcommand{\arraystretch}{1.3}
\begin{tabular}{cccccccccccccccccc}
\hline\hline
\multicolumn{17}{c}{NGC\,2477}\\
\cline{9-11}\\
&\multicolumn{2}{c}{Evolved}&&\multicolumn{3}{c}{$\chi$}&&\multicolumn{4}{c}{$\rm Observed+Evolved$}&&\multicolumn{4}{c}{$\rm Extrapolated+Evolved$}\\
\cline{2-3}\cline{5-7}\cline{9-12}\cline{14-17}\\
Region&N$^*$&\mevol&&$0.76-1.08$&&$1.08-1.95$&&$\rm N^*$&\mobs&$\sigma$&$\rho$&&$\rm N^*$&\mtot&$\sigma$&$\rho$\\
(pc)&(Stars)&($10^2\ms$)&& && &&($10^3$stars)&($10^3\ms$)& & && ($10^3$stars)&
($10^3\ms$)& & \\
 (1)&  (2) & (3)   && (4)   &&(5)  &&(6)  &(7) & (8)  & (9)  && (10)&(11)&(12)&(13)\\
\hline
$0.0-1.4$&$30\pm6$&$0.6\pm0.1$&&$-2.3\pm0.5$&&$0.0\pm0.2$&&$0.3\pm0.03$&$0.4\pm0.04$&$66\pm7$&$37\pm4$&&$0.7\pm0.3$&$0.5\pm0.1$&$85\pm13$&$45\pm7$&\\
$1.4-5.2$&$81\pm8$&$1.6\pm0.2$&&$-0.6\pm0.4$&&$1.4\pm0.2$&&$1.2\pm0.07$&$1.5\pm0.1$&$20\pm1$&$2.8\pm0.2$&&$5.4\pm3.3$&$2.7\pm0.6$&$34\pm8$&$4.6\pm1.1$&\\
$5.2-11.6$&$19\pm9$&$0.4\pm0.2$&&$1.9\pm0.7$&&$3.3\pm0.4$&&$0.7\pm0.07$&$0.7\pm0.1$&$2.1\pm0.2$&$0.13\pm0.01$&&$6.2\pm4.4$&$2.2\pm0.9$&$7\pm3$&$0.4\pm0.1$&\\
$0.0-11.6$&$128\pm12$&$2.5\pm0.2$&&$0.4\pm0.4$&&$1.5\pm0.1$&&$2.2\pm0.11$&$2.7\pm0.1$&$6.3\pm0.3$&$0.43\pm0.02$&&$12.1\pm7.7$&$5.3\pm1.5$&$13\pm4$&$0.8\pm0.2$&\\
\hline
\multicolumn{17}{c}{NGC\,2516}\\
\cline{9-11}\\
&\multicolumn{2}{c}{Evolved}&&\multicolumn{3}{c}{$\chi$}&&\multicolumn{4}{c}{$\rm Observed+Evolved$}&&\multicolumn{4}{c}{$\rm Extrapolated+Evolved$}\\
\cline{2-3}\cline{5-7}\cline{9-12}\cline{14-17}\\
Region&N$^*$&\mevol&&$0.39-0.90$&&$0.90-4.17$&&$\rm N^*$&\mobs&$\sigma$&$\rho$&&$\rm N^*$&\mtot&$\sigma$&$\rho$\\
(pc)&(Stars)&($10^2\ms$)&& && &&($10^2$stars)&($10^2\ms$)& & && ($10^2$stars)&
($10^2\ms$)& & \\
\hline
$0.0-0.6$&---&---&&$-1.1\pm0.5$&&$0.7\pm0.2$&&$0.6\pm0.1$&$0.9\pm0.1$&$79\pm14$&$93\pm16$&&$0.8\pm0.2$&$0.9\pm0.2$&$82\pm14$&$103\pm17$&\\
$0.6-1.7$&---&---&&$0.3\pm0.3$&&$0.7\pm0.2$&&$1.6\pm0.2$&$1.9\pm0.3$&$26\pm4$&$11\pm2$&&$3.9\pm1.2$&$2.4\pm0.4$&$32\pm5$&$13.3\pm2.1$&\\
$1.7-6.2$&---&---&&$0.5\pm0.3$&&$1.8\pm0.1$&&$7.4\pm1.2$&$6.4\pm0.8$&$5.5\pm0.6$&$0.6\pm0.1$&&$26.9\pm10.3$&$9.8\pm1.8$&$8.6\pm1.6$&$1.0\pm0.2$&\\
$0.0-6.2$&---&---&&$0.4\pm0.2$&&$1.4\pm0.1$&&$9.6\pm1.1$&$9.1\pm0.8$&$7.4\pm0.7$&$0.9\pm0.1$&&$30.1\pm8.7$&$12.8\pm1.6$&$10.5\pm1.3$&$1.3\pm0.2$&\\
\hline\hline
\end{tabular}
\begin{list}{Table Notes.}
\item Cols.~4 and 5 give the MF slopes derived for the low-mass and high-mass MS ranges. Notice that the 
low-mass MS ranges are $\rm 0.76\leq m(\ms)\leq 1.08$\ and $\rm 0.39\leq m(\ms)\leq 90$, respectively for
NGC\,2477 and NGC\,2416. The high-mass MS ranges are $\rm 1.08\leq m(\ms)\leq 1.50$\ and $\rm 0.90\leq 
m(\ms)\leq 4.17$. The mass of the evolved  stars is included in \mobs\ (col.~7) and \mtot\ (col.~11).
Units of $\sigma$ (cols.~8 and 12) and $\rho$ (cols.~9 and 13) are $\rm \ms\,pc^{-2}$ and $\rm 
\ms\,pc^{-3}$, respectively.
\end{list}
\end{scriptsize}
\end{table*}

\subsubsection{Mass segregation}
\label{Mass_segr}

Within the uncertainty, the overall MF slope ($\chi=1.5\pm0.1$) in the mass range $\rm 1.08\leq 
m(\ms)\leq 1.95$ is similar to a standard Salpeter ($\chi=1.35$) IMF. However, the MF slope presents 
large variations in the inner regions, being flat ($\chi=0.0\pm0.2$) in the core and steep 
($\chi=3.3\pm0.24$) in the region $\rm 5.2\leq r(pc)\leq11.6$. MF slope flattening from the outskirts 
to the core is detected in the mass range $\rm 0.76\leq m(\ms)\leq 1.08$\ as well. Similarly to the mass 
range $\rm 1.08\leq m(\ms)\leq 1.95$ the flattest MF ($\chi=-2.3\pm0.5$) occurs in the core, while the 
steepest MF ($\chi=1.9\pm0.7$) occurs in the region $\rm 5.2\leq r(pc)\leq11.6$. This fact reflects 
the advanced dynamical state of NGC\,2477, particularly the effects of large-scale mass segregation, 
in the sense that low-mass stars originally in the core are being transferred to the cluster's outskirts
while massive stars accumulate in the core. This produces a flat MF in the core and a steep one in the 
halo.

Another evidence pointing to large-scale mass segregation is the large difference 
in the number-density of giants in each cluster region with respect to that
of MS stars. We assume here as giants the stars brighter than the turnoff (with mass $\rm 
m\geq1.98$\,\ms) and as MS stars those with mass in the range $\rm 0.08\leq m(\ms)\leq1.98$. 
For the giants the number-density ratio is $\rm\frac{\rho^*(core)}{\rho^*(overall)}\sim130$, 
while for the MS stars the ratio drops to $\rm\frac{\rho^*(core)}{\rho^*(overall)}\sim30$. This 
means that in the core of NGC\,2477 the number-density of giants with respect to MS 
stars is $\sim4$ larger than in the cluster as a whole. The situation gets reversed in the region 
$\rm 5.2\leq r(pc)\leq11.6$, where the giants and MS number-density ratios are, respectively 
$\rm\frac{\rho^*(core)}{\rho^*(overall)}\sim0.15$ and $\sim0.61$. Consequently, in this region the 
number-density of MS stars with respect to the giants is $\sim4$ larger than in the 
cluster as a whole. 

Mass segregation in a star cluster scales with the relaxation time, defined as $\rm 
\tr=\frac{N}{8\ln N}\tcr$, where $\rm\tcr=R/\sigma_v$ is the crossing time, N is the (total) number 
of stars and $\rm\sigma_v$\ is the velocity dispersion (Binney \& Tremaine \cite{BinTre1987}). 
$\tr$ is the characteristic time scale in which the cluster reaches some level of kinetic 
energy equipartition with massive stars sinking to the core and low-mass stars being transferred to 
the halo. Assuming a typical $\rm\sigma_v\approx3\,\kms$ (Binney \& Merrifield \cite{Binney1998}) 
we obtain for the whole cluster $\rm\tr\sim600$\,Myr, and for the core $\rm\tr\sim6$\,Myr.
Consequently, the presence of mass segregation and thus MF slope flattening, particularly in the 
core, is consistent with both estimates of $\tr$ which are smaller than the age of NGC\,2477.

\subsection{The moderately young open cluster NGC\,2516}
\label{n2516}

\begin{figure}
\resizebox{\hsize}{!}{\includegraphics{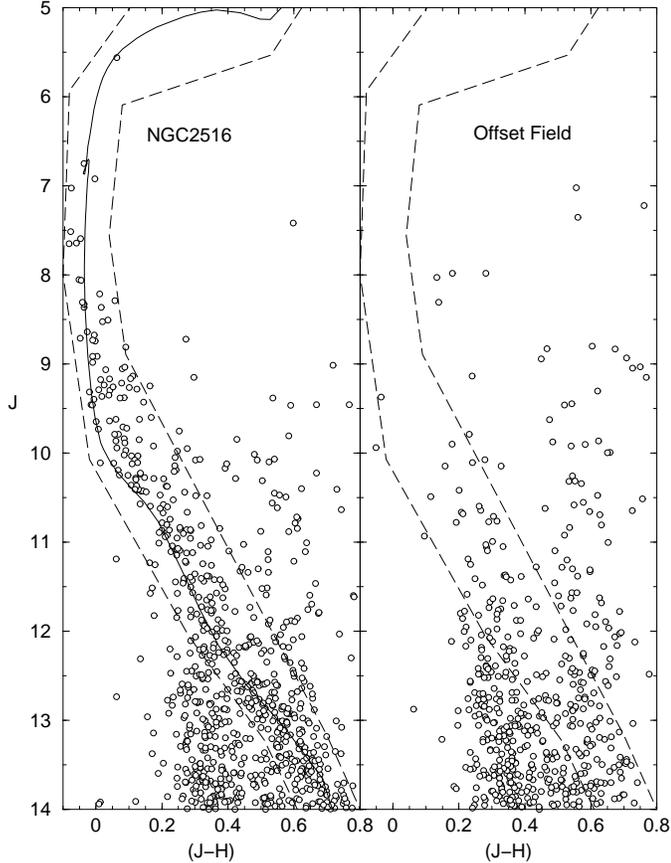}}
\caption[]{Left panel: $\jj\times\jh$ CMD of the central 20\arcmin\ ($\sim2.5$\,pc) of 
NGC\,2516 with the {\em best-fit} Padova isochrone ($\rm age\sim160$\,Myr) superimposed. 
Right panel: CMD of the corresponding (same area) offset field. The colour-magnitude filter 
is shown in both panels. Note that the MS of NGC\,2516 is less populated than that of NGC\,2477
in Fig~\ref{fig1}.}
\label{fig4}
\end{figure}

The analysis of NGC\,2516 followed the same steps as that of NGC\,2477, noting that in the case 
of NGC\,2516 the WEBDA coordinates (Table~\ref{tab1}) corresponded to the actual cluster center. 
The Galactic coordinates of NGC\,2516 are $\ell=273.82^\circ$ and $b=-15.86^\circ$.
Because of the large projected area of NGC\,2516 we used a radius of $\rm r=80\arcmin$ to extract
the 2MASS photometry. The $\jj\times\jh$ CMD of the central ($\rm r=20\arcmin\sim2.5$\,pc) region
of NGC\,2516 is shown in the left panel of Fig.~\ref{fig4}, and the corresponding, same area 
(extracted from $\rm77.46\arcmin\leq r\leq80\arcmin$) offset field is in the right 
panel. The young age of NGC\,2516 is apparent in the CMD morphology, particularly the extended, 
nearly vertical MS and absence of evolved stars. 

\subsubsection{Structural parameters}

The {\em best-fit} to the CMD of NGC\,2516 was obtained with the 160\,Myr Padova isochrone, a 
colour excess $\ejh=0.0$ ($\ebv=0.0$), a distance modulus $\mMJ=8.2\pm0.1$, and a distance from 
the Sun $\ds=0.44\pm0.02$\,kpc. With this value the Galactocentric distance of NGC\,2516 turns 
out to be $\dgc=8.0\pm0.1$\,kpc. Considering the uncertainties associated to the isochrone fit 
the age of NGC\,2516 gets constrained to the range $160\pm10$\,Myr. This value agrees with that
estimated by Jeffries, Thurston \& Hambly (\cite{Jeff2001}). 

\begin{figure}
\resizebox{\hsize}{!}{\includegraphics{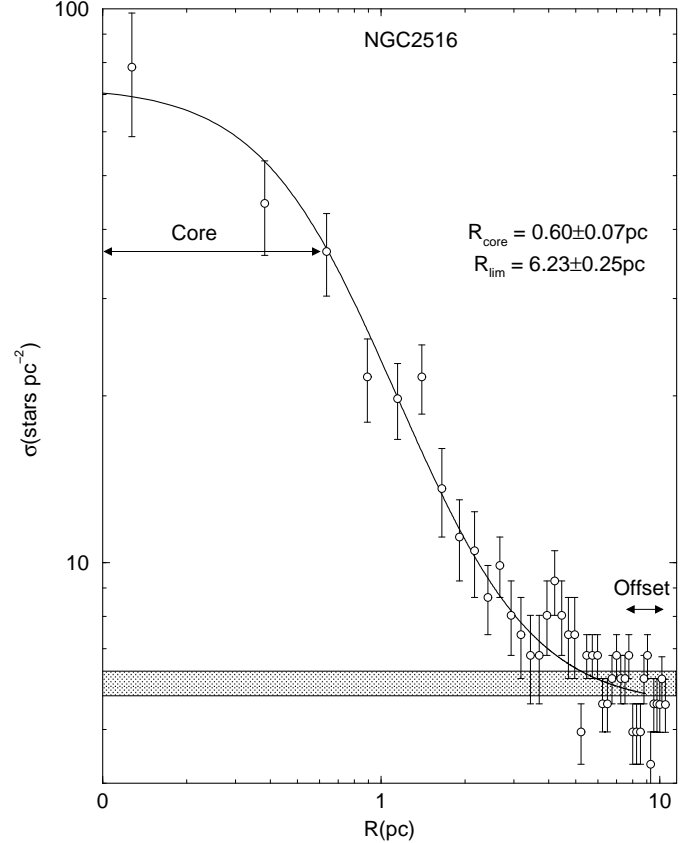}}
\caption[]{Same as Fig.~\ref{fig2} for the radial density profile of NGC\,2516.}
\label{fig5}
\end{figure}

The radial density profile of stars in NGC\,2516 (after applying the colour-magnitude filter - 
Fig.~\ref{fig4}) is shown in Fig.~\ref{fig5}, where the background contribution corresponds to 
the average density of stars in the region $\rm7.6\leq r(pc)\leq10.18$. Similarly to NGC\,2477 
we restricted the radial density analysis to filtered stars with $\jj\leq13.0$, corresponding to 
$\rm M_J\leq4.8$ and mass $\rm m\geq0.8\,\ms$ (Sect.~\ref{MF2516}). At $\jj=13.0$ the photometric 
uncertainties amount to $\rm\epsilon_J\approx0.03$ (Bonatto, Bica \& Santos Jr. \cite{BBS2005}).
The fit with the two-parameter King (\cite{King1966a}) surface density profile to the background-subtracted
radial density profile resulted in $\rc=0.6\pm0.1$\,pc and a projected central density of stars 
of $\rm\sigma_0\approx67\pm9\,stars\,pc^{-2}$. We derived a limiting radius of $\rl=6.2\pm0.2$\,pc, 
a value $\sim3$\ times larger than the tidal radius estimated by Bergond, Leon \& Guibert 
(\cite{Bergond2001}) and Tadross et al. (\cite{Tad2002}). We note that in the present work
the tidal radius could not be calculated since the nonlinear least-squares routine did not 
reach convergence with the three-parameter King (\cite{King1962}) model applied to the 
background-subtracted radial distribution of stars in NGC\,2516 (Fig.~\ref{fig5}). 

\subsubsection{Mass functions}
\label{MF2516}

\begin{figure}
\resizebox{\hsize}{!}{\includegraphics{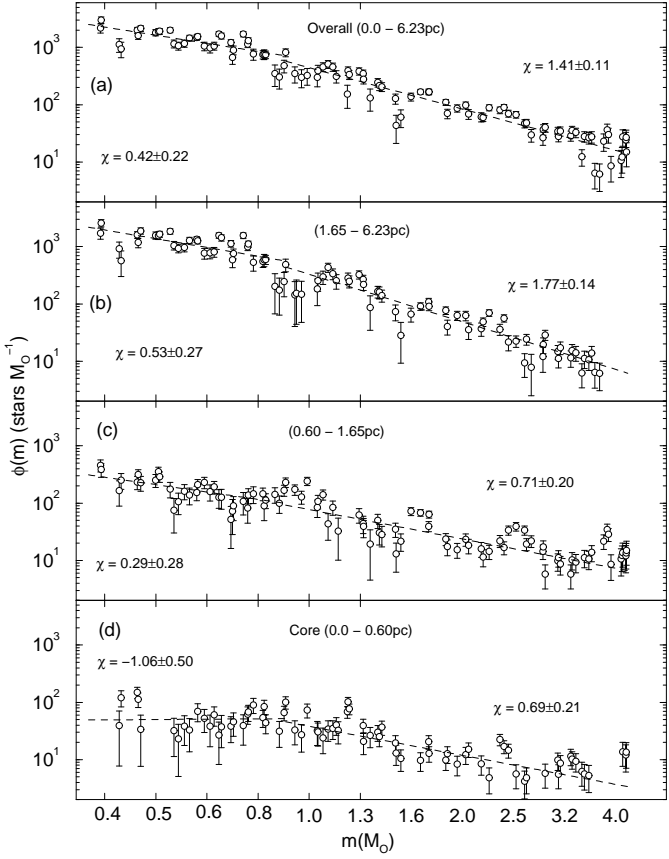}}
\caption[]{Mass functions in different spatial regions of NGC\,2516. Each panel contains MFs derived
independently from the \jj, \hh\ and \ks\ 2MASS photometry. A break in the MFs at $\rm\mb\approx0.9$\,\ms\ 
can be seen particularly in the core region (panel (d)). MF fits ($\phi(m)\propto m^{-(1+\chi)}$) are 
shown as dashed lines.}
\label{fig6}
\end{figure}

Similarly to NGC\,2477 we derive MFs for the following internal regions of NGC\,2516: {\em (i)} $\rm 
0.0\leq r(pc)\leq0.6$ (core), {\em (ii)} $\rm 0.6\leq r(pc)\leq1.7$, {\em (iii)} $\rm 1.7\leq r(pc)
\leq6.2$\ and {\em (iv)} $\rm 0.0\leq r(pc)\leq6.2$ (overall). The colour-magnitude filter 
(Fig.~\ref{fig4}) has been applied to the CMDs of the above regions and offset field
($\rm7.6\leq r(pc)\leq10.18$) to take into account the Galactic contamination. The resulting MFs 
covering the mass range $\rm 0.39\leq m(\ms)\leq4.17$ are shown in Fig.~\ref{fig6}.

Parameters derived from the MFs, such as observed and extrapolated mass, projected and mass 
densities in each region of NGC\,2516 are given in Table~\ref{tab2}. Similarly to NGC\,2477
a break in the MFs can be seen, particularly in the core region at $\rm m=0.9\,\ms$. In addition, 
mass-segregation effects are reflected on the slope variation with distance to the center of NGC\,2516. 
In the mass range $\rm 0.90\leq m(\ms)\leq4.17$\ the MF slope is similar ($\chi\approx0.7$) both 
for the core and the region $\rm 0.6\leq r(pc)\leq1.7$, increasing to $\chi\approx1.8$ for
$\rm 1.7\leq r(pc)\leq6.2$. In this mass range the overall MF is similar to a standard 
Salpeter law. In the range $\rm 0.39\leq m(\ms)\leq0.90$\ the core MF is very flat 
($\chi\approx-1.1$) while in the other regions the slope is similar to that of the overall
MF ($\chi\approx0.4$).

The sharp flattening in the MFs of NGC\,2516 was previously detected by Jeffries, Thurston \& 
Hambly (\cite{Jeff2001}) based on CCD photometry of the spatial region internal to 
$\rm r\approx31\arcmin\ (\approx4\,pc)$. In the range $\rm 0.3\leq m(\ms)\leq0.7$ they derive MF 
slopes of $\chi=-0.75\pm0.20$\ or $\chi=-1.00\pm0.18$, depending on the (solar metallicity) isochrone 
models used. In the range $\rm 0.7\leq m(\ms) \leq3.0$ they derive slopes $\chi=1.47\pm0.11$ or 
$\chi=1.58\pm0.11$, for the same models as above. Considering the slope uncertainties and differences 
in the spatial regions sampled, the MF slopes derived in the present work are consistent with those 
of Jeffries, Thurston \& Hambly (\cite{Jeff2001}), in both mass ranges. In addition, the total mass 
derived in the present work $\mtot=1280\pm162$\,\ms\ agrees with the value estimated by Jeffries, 
Thurston \& Hambly (\cite{Jeff2001}), $\rm m\approx 1240-1560$\,\ms\ (corrected for mass segregation
for stars with $\rm m\geq0.3\,\ms$). We note that previous works gave exceedingly low values for 
the stellar mass in NGC\,2516, e.g. Bergond, Leon \& Guibert (\cite{Bergond2001}) estimate 
$\mtot\approx170$\,\ms\ (for stars with $\rm m\geq1.1\,\ms$), while in Tadross et al. 
(\cite{Tad2002})$, \mtot\approx36\,\ms$. Extrapolated and observed core mass agree, while the
extrapolated overall mass is $\sim40\%$ larger than the mass locked up in the observed stars.

Mass segregation in NGC\,2516, as implied by the spatial variation of the MF slopes (Fig.~\ref{fig6}),
is consistent with the overall $\rm\tr\sim90$\,Myr and $\rm\tr\sim0.4$\,Myr in the
core, both estimates are smaller than the age of the cluster.

\subsection{Comparing dynamical states}
\label{dyna}

The less-advanced dynamical state of NGC\,2516 as compared to that of NGC\,2477 is reflected both in 
the absolute value of the MF slopes and in the spatial rate of change of $\chi$. In the mass range 
$\rm 0.90\leq m(\ms)\leq4.17$\ of NGC\,2516 the MF slope varies from $\chi\approx0.7$\ in the core 
to $\chi\approx1.8$\ in the region $\rm 1.7\leq r(pc)\leq6.2$, while in the equivalent mass range 
$\rm1.08\leq m(\ms)\leq1.95$\ of NGC\,2477 the slope varies from the flatter $\chi\approx0.0$\ in the 
core to the steeper $\chi\approx3.3$\ in the region at $\rm5.2\leq r(pc)\leq11.6$. Thus, in this mass 
range the radial variation of the MF slope in NGC\,2516 is $\rm\frac{\Delta\chi}{\Delta r}=(0.30
\pm0.19)\,pc^{-1}$, while in NGC\,2477 $\rm\frac{\Delta\chi}{\Delta r}=(0.43\pm0.19)\,pc^{-1}$. A similar 
situation with respect to MF slope values and radial variations occurs in the mass range corresponding 
to the MF breaks in NGC\,2516 with $\rm\frac{\Delta\chi}{\Delta r}=(0.4\pm0.3)\,pc^{-1}$ and 
NC\,2477 with $\rm\frac{\Delta\chi}{\Delta r}=(0.5\pm0.2)\,pc^{-1}$. Finally, the overall MF 
slopes in NGC\,2516 and NGC\,2477 are essentially the same in the low-mass MS ($\chi\approx0.4$) and 
in the high-mass MS ($\chi\approx1.4$). 

Thus, the discussions above show that the fact that NGC\,2477 is $\sim7$ times as old as NGC\,2516 
turns out to be more important to the internal dynamical evolution than NGC\,2477 being $\sim4$ times 
more massive than NGC\,2516.

\section{Multi-parametric analysis}
\label{MPA}

The analysis method described in Sects.~\ref{n2477} and \ref{n2516} respectively for NGC\,2477 and 
NGC\,2516 was applied to the remaining clusters in Table~\ref{tab1}. We derived a homogeneous set 
of parameters associated to the internal structure (core and limiting radii), stellar evolution 
(extrapolated mass) and dynamical evolution (core and overall mass and mass density, and MF slope). 
We do not include in the statistical analyses below the tidal radius, since convergence in the 
nonlinear least-squares routine with the three-parameter King (\cite{King1962}) model has been 
reached only for the background-subtracted radial profiles of M\,26 
($\rm\rt=15\pm7\,pc\sim1.9\times\rl$), NGC\,2477 ($\rm\rt=16\pm2\,pc\sim1.4\times\rl$), M\,67 
($\rm\rt=16\pm3\,pc\sim1.4\times\rl$) and NGC\,188 ($\rm\rt=22\pm3\,pc\sim1.8\times\rl$). 
As a comparison between conceptually different estimates of cluster dimension, in these 
four open clusters the ratio $\rt/\rl$\ varies from 1.4 to 1.9. Thus, for sample completeness
purposes we use the limiting radius in what follows.

Except for cluster age and Galactocentric distance which are listed in Table~\ref{tab1}, the 
remaining parameters are listed in Table~\ref{tab3}, in which we consider separately the core and 
overall regions. By columns: (1) - cluster designation; (2) - core radius; (3) - observed core mass, 
which includes the mass of the evolved stars down to the mass corresponding to the 2MASS faint 
magnitude limits (Sect.~\ref{MF}); (4) - extrapolated core mass, including stars with $\rm m=0.08$\,\ms; 
(5) - projected core mass density ($\rm m/\pi R^2_{core}$); (6) - core mass density 
($\rm m/\frac{4}{3}\pi R^3_{core}$); (7) - core MF slope (for the mass range $\rm m\geq\mb$); (8) - 
limiting radius; (9) - observed 
overall mass; (10) - extrapolated overall mass; (11) - overall projected mass density; and (12) - 
overall mass density. The overall MF slope ($\rm m\geq\mb$) is given in col.~13.
In col.~14 we give \mb, the stellar mass where the MF break occurs.  Finally, in col.~15 we give 
$\rm c^*=\log\left(\rl/\rc\right)$, which can be measured essentially in all open clusters. Derived 
from the limiting radius, $c^*$\ is not the same as the concentration parameter of globular clusters, 
which is based on the tidal radius (Trager, King \& Djorgovski \cite{TKD95}). Since the tidal radius 
can only be measured for the most populous open clusters and/or those at high Galactic latitudes (see
above), we emphasize the importance of using in the future $c^*$\ for statistical purposes when dealing 
with samples of open clusters. The relatively narrow range of $c^*$\, from $\approx0.8$\ to $\approx1.0$\ 
is characteristic of loose star clusters (Trager, King \& Djorgovski \cite{TKD95}).

\begin{table*}
\caption[]{Structural and dynamical-related parameters of the target open clusters.}
\begin{scriptsize}
\label{tab3}
\renewcommand{\tabcolsep}{0.40mm}
\renewcommand{\arraystretch}{1.3}
\begin{tabular}{lcccccccccccccccc}
\hline\hline
&&\multicolumn{6}{c}{Core}&&\multicolumn{6}{c}{Overall}\\
\cline{2-8}\cline{10-15}\\
Cluster&&\rc&\mobs&\mtot&$\sigma$&$\rho$&$\chi$&&\rl&\mobs&\mtot&$\sigma$&$\rho$&$\chi$&$\mb$&$c^*$\\
      &&(pc)&($\rm 10^2\ms$)&($\rm 10^2\ms$)& & &&&
      (pc)&($\rm 10^2\ms$)&($\rm 10^2\ms$)& & &&(\ms)\\
    (1)&&(2)&(3)&(4)&(5)&(6)&(7)&&(8)&(9)&(10)&(11)&(12)&(13)&(14)&(15)\\
\hline
M\,26    &&$0.8\pm0.1$&$1.2\pm0.3$&$2.9\pm0.9$&$150\pm48$&$142\pm46$&$1.3\pm0.2$&&
         $8.0\pm0.5$&$6.8\pm1.6$&$19\pm7$&$9.4\pm3.3$&$0.9\pm0.3$&$1.5\pm0.2$&(\dag)&1.0\\

NGC\,2516&&$0.6\pm0.1$&$0.9\pm0.1$&$0.9\pm0.2$&$82\pm14$&$103\pm17$&$0.7\pm0.2$&&
         $6.2\pm0.2$&$9.1\pm0.8$&$13\pm2$&$10.5\pm1.3$&$1.3\pm0.2$&$1.4\pm0.1$&0.90&1.0\\

NGC\,2287&&$1.1\pm0.1$&$0.8\pm0.2$&$0.9\pm0.2$&$22\pm5$&$15\pm4$&$0.2\pm0.3$&&
         $7.0\pm0.7$&$5.7\pm0.7$&$7.2\pm0.9$&$4.7\pm0.6$&$0.5\pm0.1$&$1.5\pm0.2$&1.01&0.8\\

M\,48    &&$0.9\pm0.2$&$0.5\pm0.1$&$0.6\pm0.1$&$26\pm3$&$23\pm3$&$0.3\pm0.2$&&
         $8.8\pm0.4$&$4.8\pm0.7$&$6.3\pm1.5$&$2.6\pm0.6$&$0.22\pm0.05$&$1.3\pm0.2$&1.06&1.0\\

M\,93    &&$0.7\pm0.1$&$1.1\pm0.1$&$1.4\pm0.1$&$85\pm8$&$88\pm9$&$-0.2\pm0.2$&&
         $7.2\pm0.6$&$7.4\pm0.6$&$17\pm5$&$10.2\pm3.4$&$1.0\pm0.3$&$1.2\pm0.1$&---&1.0\\

NGC\,5822&&$1.1\pm0.1$&$1.1\pm0.1$&$1.6\pm0.4$&$43\pm11$&$30\pm7$&$0.6\pm0.2$&&
         $8.0\pm0.4$&$13.6\pm1.0$&$26\pm10$&$13\pm5$&$1.2\pm0.4$&$1.7\pm0.1$&---&0.9\\

NGC\,2477&&$1.4\pm0.1$&$4.1\pm0.4$&$5.2\pm0.8$&$85\pm13$&$45\pm7$&$0.0\pm0.2$&&
         $11.6\pm0.7$&$27\pm14$&$53\pm16$&$13\pm4$&$0.8\pm0.2$&$1.5\pm0.1$&1.08&0.9\\

NGC\,3680&&$0.5\pm0.2$&$0.3\pm0.1$&$0.4\pm0.1$&$45\pm6$&$64\pm9$&$-1.1\pm0.4$&&
         $5.8\pm0.6$&$1.7\pm0.2$&$3.6\pm1.1$&$3.4\pm1.0$&$0.5\pm0.1$&$0.5\pm0.3$&---&1.0\\

IC\,4651 &&$0.8\pm0.1$&$0.9\pm0.1$&$1.0\pm0.1$&$54\pm6$&$53\pm6$&$-1.4\pm0.3$&&
         $6.0\pm0.3$&$7.3\pm0.7$&$8.4\pm1.0$&$7.5\pm0.9$&$0.9\pm0.1$&$0.6\pm0.3$&0.97&0.9\\

M\,67    &&$1.1\pm0.1$&$1.3\pm0.2$&$1.4\pm0.2$&$34\pm4$&$22\pm3$&$-2.5\pm0.4$&&
         $11.7\pm0.6$&$9.4\pm0.7$&$9.9\pm1.2$&$2.5\pm0.3$&$0.16\pm0.02$&$-0.3\pm0.2$&0.80&1.0\\
         
NGC\,188&&$1.3\pm0.1$&$0.7\pm0.1$&$6.1\pm0.3$&$99\pm42$&$53\pm23$&$0.6\pm0.7$&&
         $11.9\pm0.5$&$3.8\pm0.1$&$38\pm16$&$8.5\pm3.6$&$0.54\pm0.23$&$1.9\pm0.7$&(\ddag)&1.0\\
\hline
\end{tabular}
\begin{list}{Table Notes.}
\item  \mobs\ and \mtot\ in cols.~3, 4, 9 and 10 includes the mass in the evolved stars mass. \mb\ in
col.~14 is the mass where the MF break begins. $\chi$\ in cols.~7 and 13 is the MF slope in the mass range $\rm 
m\geq\mb$. (\dag) - because of the distance from the Sun, the low-mass end of the MF of M\,26 occurs at 
$\rm m=0.98$\,\ms. (\ddag) - because of the distance from the Sun and old age the MF of NGC\,188 is 
restricted to the mass range $0.9\,\ms - 1.0\,\ms$. The uncertainty in \mb\ in col.~14 is
$\approx0.05\,\ms$ in all clusters. The uncertainty in $c^*$\ in col.~15 is 0.1 in all clusters 
except for NGC\,3680 for which it is 0.2. Units of $\sigma$ (cols.~5 and 11) and $\rho$ (cols.~6 and 12)
are  $\rm \ms\,pc^{-2}$ and $\rm \ms\,pc^{-3}$, respectively.
\end{list}
\end{scriptsize}
\end{table*}

To characterize the sample we present average values of some parameters. For the massive
($\rm m\geq1\,000$\,\ms) clusters we obtain $\langle\rc\rangle=1.0\pm0.3$\,pc, 
$\rm\langle M_{core}\rangle=(3\pm2)\times10^2\,\ms$, 
$\rm\langle\chi_{core}\rangle=0.5\pm0.5$, $\langle\rl\rangle=8.8\pm2.3$\,pc, 
$\rm\langle M_{overall}\rangle=(2.8\pm1.5)\times10^3\,\ms$ and 
$\rm\langle\chi_{overall}\rangle=1.5\pm0.2$.
For the less-massive clusters they are $\langle\rc\rangle=0.9\pm0.2$\,pc, 
$\rm\langle M_{core}\rangle=(0.8\pm0.4)\times10^2\,\ms$, 
$\rm\langle\chi_{core}\rangle=-0.9\pm1.1$, $\langle\rl\rangle=7.9\pm2.5$\,pc, 
$\rm\langle M_{overall}\rangle=(0.7\pm0.2)\times10^3\,\ms$ and 
$\rm\langle\chi_{overall}\rangle=0.7\pm0.7$. On average the core and overall MFs of the less-massive
clusters tend to be flatter than those of the massive ones. Core and overall masses of the
massive clusters are $\sim4$ times larger than those of the less-massive ones. Within the standard 
deviations core and limiting radii are similar on average, in both types of clusters.

Because of the reduced number of clusters in this work, the results below should be taken as a 
sampling of the class of open clusters. On the other hand, the statistical significance and 
homogeneous method from which the parameters were derived guarantee that the results may serve 
as indicative of trends and correlations. 

\subsection{Cluster structural and evolutionary parameters {\em vs.} cluster age}
\label{cpca}

\begin{figure} 
\resizebox{\hsize}{!}{\includegraphics{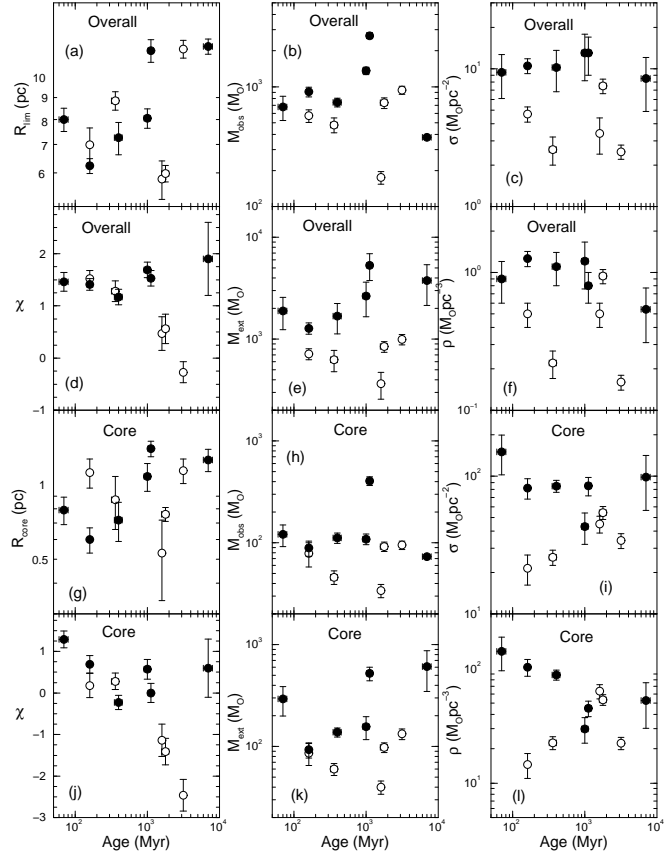}}
\caption[]{Relations of cluster parameters with cluster age. Filled symbols correspond to massive
clusters (overall mass $\geq1\,000$\,\ms).}
\label{fig7}
\end{figure}

Relations involving cluster parameters and cluster age are investigated in Fig.~\ref{fig7}, in
which we use different symbols to identify massive and less-massive 
clusters. Among the parameters tested the best correlation with age occurs for \rl\ (panel 
(a)). This correlation should be expected since young ($\rm age\leq700$\,Myr, Friel \cite{Friel95}) 
clusters with large diameters probably are not bound systems (Lyng\aa\ \cite{Lynga82}, Janes, Tilley 
\& Lyng\aa\ \cite{JTL88}), and small, massive clusters will be dissolved due to the strong effects of 
the internal dynamics (Tadross et al. \cite{Tad2002}). This correlation was not present in 
the sample studied by Tadross et al. (\cite{Tad2002}). The core radius in panel (g) does
not seem to correlate with age.

An interesting pattern occurs in the plots $\rm\rho_{core}\times age$\ (panel (l)) and 
$\rm\sigma_{core}\times age$\ (panel (i)) which may reflect different evolutionary processes in the 
core of more-massive clusters with respect to less-massive ones. A similar pattern occurs in the 
plot $\rm\chi_{overall}\times age$\ (panel (d)), in which massive and less-massive clusters seem 
to evolve differently, in the sense that old, less-massive clusters tend to have flatter overall MFs 
than old massive ones. Because of their looser nature with respect to massive clusters, old
less-massive clusters have lost a significant fraction of the low-mass content through internal 
processes such as mass segregation and evaporation, and external ones such as tidal stripping,
disk shocking and encounters with molecular clouds (Bergond, Leon \& Guibert \cite{Bergond2001}). 
As a consequence, a higher flattening degree in the MFs of old less-massive clusters with respect 
to old massive ones, is expected. A similar pattern occurs with respect to the 
relation of the core MF slope with age in panel (j).

The projected overall mass density (panel (c)), total core mass (panel (k)) and overall mass density
panel (f)) do not correlate with age, but massive and less-massive clusters in general occupy 
distinct positions in the plots.

Considering the significant scatter in the plots we find no indication of correlation with cluster 
age of total overall mass (panel (e)), observed overall mass (panel (b)) and observed core mass 
(panel (h)). A larger cluster sample is necessary to check these points.

\subsection{Cluster structural and evolutionary parameters {\em vs.} Galactocentric distance}
\label{cpgd}

\begin{figure} 
\resizebox{\hsize}{!}{\includegraphics{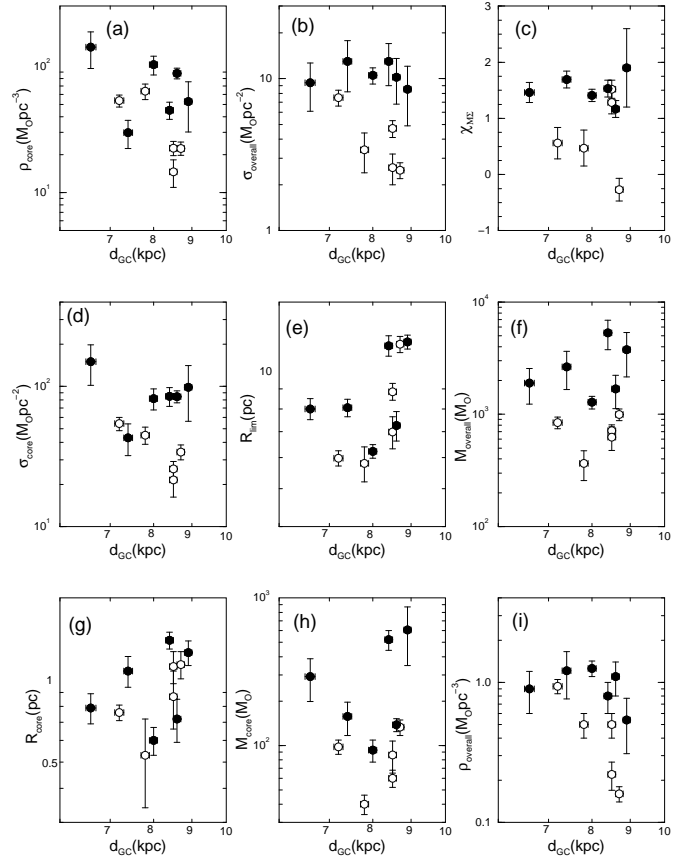}}
\caption[]{Relations of cluster parameters with Galactocentric distance. Symbols as in
Fig.~\ref{fig7}.}
\label{fig8}
\end{figure}

Relations involving cluster parameters and Galactocentric distance are investigated in 
Fig.~\ref{fig8}. The plot in panel (f) suggests that there is no correlation of cluster mass
with Galactocentric distance. Although the scatter in most plots is considerable, the plot in 
panel (e) suggests a correlation of \rl\ with Galactocentric distance \dgc, similar to the results 
of Lyng\aa\ (\cite{Lynga82}) and Tadross et al. (\cite{Tad2002}). This correlation holds for 
massive and less-massive clusters. The core mass density (panel (a)) and, to a less extent the 
overall mass density (panel (i)), seem to fall for clusters at larger \dgc, 
probably because of the slight dependence of core radius on \dgc\ (panel (g)). 
Finally, in panel (i) of Fig.~\ref{fig9} we examine the relation of \dgc\ with
cluster age. Young and old clusters, with any mass, are similarly
distributed in terms of \dgc. This apparently contradicts the conclusion of 
Lyng\aa\ (\cite{Lynga82}) in which older clusters are found preferentially in the outer parts of
the Galaxy, while younger clusters are evenly distributed. However, our result
is based on 11 clusters only, which sample a limited range in \dgc\ in the
solar surroundings.  

\subsection{Core {\em vs.} overall parameters}
\label{cop}

\begin{figure} 
\resizebox{\hsize}{!}{\includegraphics{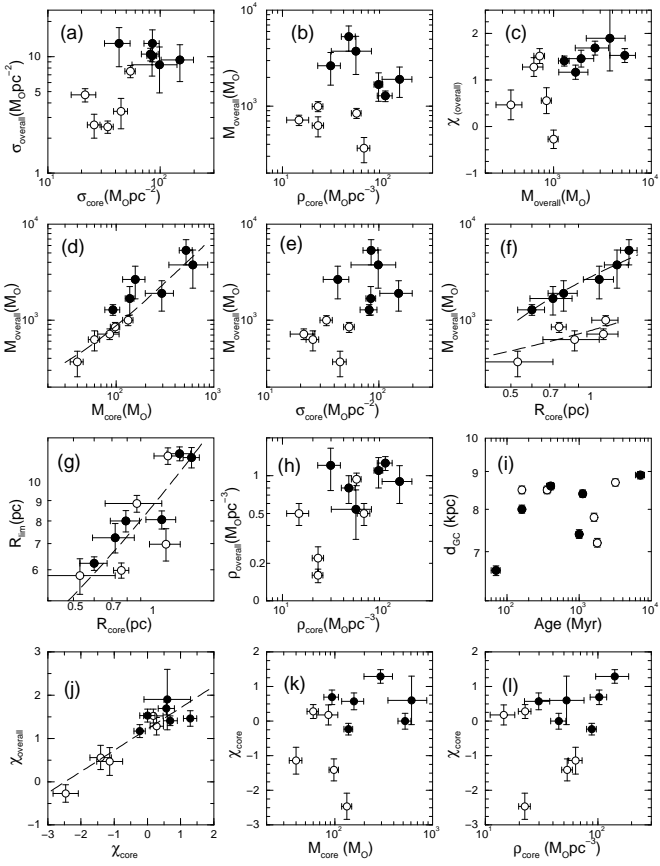}}
\caption[]{Relations involving core and overall parameters. Symbols as in
Fig.~\ref{fig7}. Dashed lines correspond to linear least-squares fits to the points.}
\label{fig9}
\end{figure}

In Fig.~\ref{fig9} we examine relations involving core and overall parameters. The tight 
correlations involving core and limiting radii (panel (g)) and core and overall total mass (panel 
(d)) are probably consequences of size and mass scaling, in the sense that bigger clusters tend to 
have bigger cores and larger masses. Linear least-squares fits to the points resulted in correlation 
coefficients of 0.85 and 0.84, respectively for the plots in panels (d) and (g). The projected overall 
density (panel (a)) and overall mass density (panel (h)) seem to correlate as well, although with 
significant scatter. Core radius and overall mass are tightly correlated (panel (f)) as well. However, 
massive and less-massive clusters seem to follow separate, but nearly parallel, paths in the plot. 
Correlation coefficients are 0.97 and 0.68, respectively for the fits of the massive and less-massive 
clusters.

Total overall mass and overall MF slope correlate (panel (c)). In this case the MS of massive 
clusters is characterized by a narrow range of MF slopes, $1\leq\chi\leq2$, while the MS of 
less-massive clusters has $\chi$\ spread over a large range. No definite pattern emerges from 
the relations of total overall mass with core mass density (panel (b)) and core projected density 
(panel (e)).

Finally, the core and overall MF slopes are tightly correlated (panel (j)), with a 
correlation coefficient of 0.90. Total core mass and core MF slope (panel (k)) present a similar
pattern as total overall mass and overall MF slope (panel (c)). 

\section{MF break and the universality of the IMF}
\label{MFT}

Since the early work of Salpeter (\cite{Salpeter55}) in which the IMF of solar-neighbourhood stars 
in the range 0.4\ms\ to 10\ms\ was described by the function $\phi(m)\propto m^{-(1+\chi)}$, with 
$\chi=1.35$, evidence has been accumulating pointing to the presence of breaks in the MF of low-mass 
stars, in the sense that it tends to be flatter than the MF of high-mass stars. On the high-mass end, 
the Salpeter IMF has been shown to extend up to very-massive stars, such as $\rm m\approx120$\,\ms\ 
(Scalo \cite{Scalo86}; Massey \cite{Massey98}). However, star counts in the solar neighbourhood 
show that the IMF for the low-mass range breaks up in two power-laws, $\chi=1.2\pm0.3$ for 
$\rm 0.5\leq m(\ms)\leq 1.0$ and $\chi=0.3\pm0.7$ for $\rm 0.08\leq m(\ms)\leq 0.5$. In the brown 
dwarf mass range the slope seems to be even flatter, $\chi\approx-0.5$ (Kroupa \cite{Kroupa2001}, 
\cite{Kroupa2002}). 

Present-day data suggest that the IMF may have a universal character, in the sense that it probably is
similar in very different environments, from star-forming regions in molecular clouds to rich star 
clusters and the field (Kroupa \cite{Kroupa2002}). As additional observational evidence of similar 
IMFs in different environments von Hippel et al. (\cite{Hippel1996}) discuss star counts in the 
near-IR, in the optical with Hubble Space Telescope data and from the Hubble Deep Field which 
indicate that field stars of the Galactic thick disk and halo have a LF similar to that of the solar 
neighbourhood down to $0.08\,\ms$ stars - stellar populations differing significantly in local
density and chemical abundances. As pointed out by Kroupa (\cite{Kroupa2002}) this scenario appears 
to indicate that the distribution of stellar masses should depend only on the process of molecular 
cloud fragmentation. In this case, the fragmentation would have to produce similar IMFs despite very 
different initial conditions, a physical process which is not yet fully understood (Kroupa 
\cite{Kroupa2002}).

\begin{figure} 
\resizebox{\hsize}{!}{\includegraphics{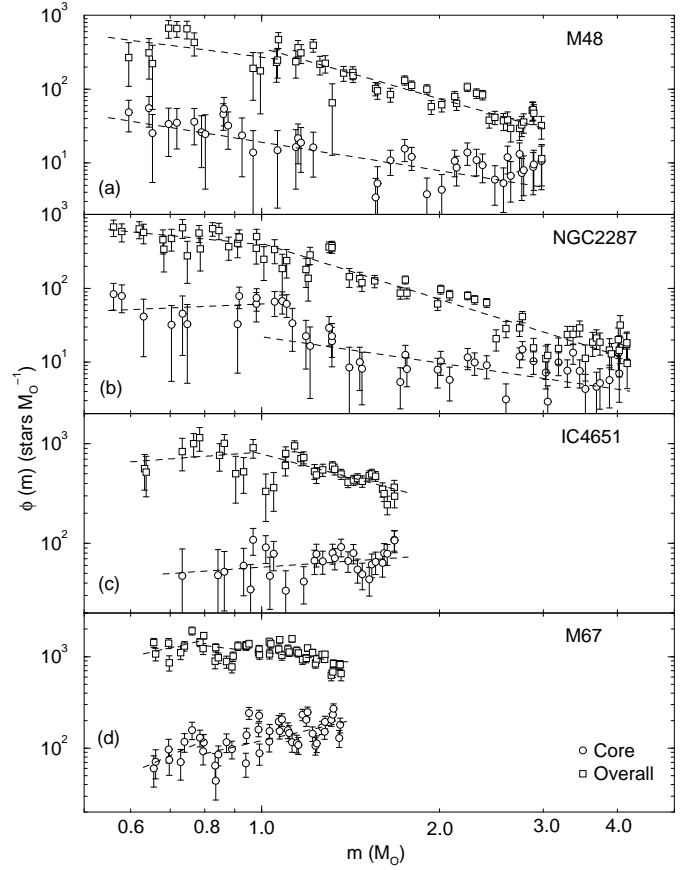}}
\caption[]{Open clusters with a MF break in the low-mass MS. Because of the significant error 
bars in the MF of NGC\,2287 (panel (b)) the low-mass MS and high-mass MS fits do not match
at $\rm m\approx1\,\ms$. Core and overall regions are considered separately.}
\label{fig10}
\end{figure}

The relatively populous nature of the open clusters included in the present study offers an  
opportunity to investigate the presence and properties of the MF break in clusters with a high 
statistical significance. However, the basically solar metallicity of the present clusters does
not allow inferences on relations of \mb\ with metallicity (e.g. von Hippel et al. \cite{Hippel1996})
or on the possible dependence of MF flattening (in the range $\rm m\leq0.7\,\ms$) with metallicity, 
as discussed in Kroupa (\cite{Kroupa2002}).

An MF slope break, followed by a MF slope flattening (turnover) for lower masses, is present in six 
clusters of the present sample. The MFs of NGC\,2477 clearly present a break at $\rm\mb\approx1.08$\,\ms\ 
(Fig.~\ref{fig3}), while in NGC\,2516 the break occurs at $\rm\mb\approx0.90$\,\ms\ (Fig.~\ref{fig6}). 
In both clusters the break consistently occurs in the four spatial MFs presented in Figs.~\ref{fig3} 
and \ref{fig6}. In the remaining clusters of the present paper the break is present in the MFs of M\,48 
at $\rm\mb\approx1.06$\,\ms\ (overall MF), NGC\,2287 at $\rm\mb\approx1.01$\,\ms\ (core and overall MF), 
IC\,4651 at $\rm\mb\approx0.97$\,\ms\ (overall MF) and M\,67 at $\rm\mb\approx0.80$\,\ms\ (core and 
overall MF). The core and overall MFs of these clusters are shown in Fig.~\ref{fig10} along with the 
respective MF fits. In col.~2 of Table~\ref{tab4} we give the mass range over which we measure the MF 
flattening ($\rm m\leq\mb$), while the corresponding core and overall MF slopes in this mass range are 
in cols.~3 and 4, respectively.

\begin{table}
\caption[]{Core and overall MF slopes for $\rm m\leq\mb$.}
\label{tab4}
\renewcommand{\tabcolsep}{3.6mm}
\begin{tabular}{lccc}
\hline\hline
Cluster&$\Delta m$&$\rm\chi(core)$&$\rm\chi(overall)$\\
&(\ms)\\
 ~~~(1)&(2)&(3)&(4)\\
\hline
NGC\,2516&$0.40-0.90$&$-1.1\pm0.5$&$+0.4\pm0.2$\\
NGC\,2287&$0.56-1.01$&$-1.3\pm0.5$&$-0.2\pm0.3$\\
M\,48    &$0.60-1.06$&$+0.3\pm0.2$&$~~0.0\pm0.7$\\
NGC\,2477&$0.76-1.08$&$-2.3\pm0.5$&$+0.4\pm0.4$\\
IC\,4651 &$0.63-0.97$&$-1.4\pm0.3$&$-1.4\pm0.7$\\
M\,67    &$0.66-0.80$&$-3.9\pm1.1$&$-2.5\pm0.2$\\    
\hline
\end{tabular}
\begin{list}{Table Notes.}
\item  Col.~2: Mass range where the MF flattening is observed. Cols.~3 and 4: Mass-function 
slope $\left(\phi(m)\propto m^{-(1+\chi)}\right)$\ for $\rm m\leq\mb$. 
\end{list}
\end{table}

\begin{figure} 
\resizebox{\hsize}{!}{\includegraphics{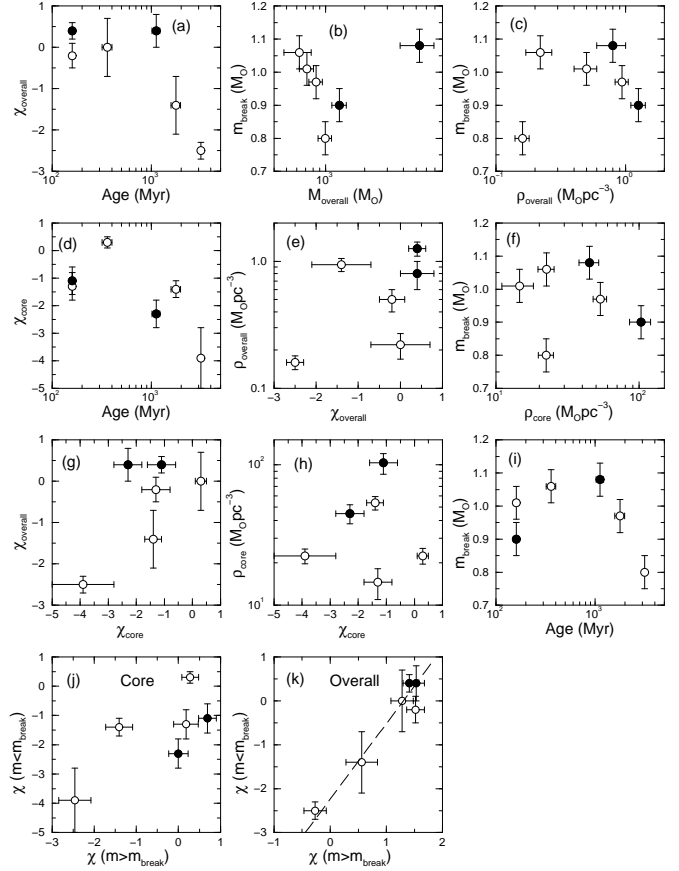}}
\caption[]{Relations involving parameters of the clusters with the MF break. $\chi_{core}$\
and $\chi_{overall}$\ in panels (a), (d), (e), (g) and (h) are the MF slopes derived for the 
range $\rm m\leq\mb$. Symbols asin Fig.~\ref{fig7}.}
\label{fig11}
\end{figure}

The flattening degree in the MF for masses below \mb\ is indeed quite high (Table~\ref{tab4}) when 
compared to that in the mass range above \mb\ (Table~\ref{tab3}), particularly in the core. Mass 
segregation throughout the cluster and low-mass star loss through dynamical evaporation certainly 
contribute to MF flattening. This effect is expected to be more significant in the old and less-massive 
clusters (Bergond, Leon \& Guibert \cite{Bergond2001}). In the mass range $\rm m\geq\mb$\ this effect 
is confirmed by the relations $\rm\chi\times age$\ both in the core and overall MFs (Fig.~\ref{fig7}, 
panels (j) and (d), respectively). In addition, observational consequences of the presence of a 
significant fraction of undetected  binaries (and higher-order multiple systems) in a cluster are, 
among others, the underestimation of the number of low-mass with respect to higher-mass stars and 
widening of the MS. Consequently, somewhat flat MFs are expected to occur in clusters with a high 
fraction of unresolved binaries (e.g. Bonatto, Bica \& Santos Jr. \cite{BBS2005} and references 
therein). Indeed, based on proper-motion considerations Bica \& Bonatto (\cite{BiBo2004}) derived 
that the fraction of undetected binaries in the cores of M\,26, NGC\,2287, M\,48, M\,93, NGC\,5822, 
NGC\,3680, IC\,4651 and M\,67 is in the range 15\% to 50\%. In the core of the old open cluster 
NGC\,188 Bonatto, Bica \& Santos Jr. (\cite{BBS2005}) estimated that the fraction of undetected binaries 
is probably around 100\%. However, even fractions of $\sim100\%$\ of unresolved binaries cannot produce 
such flat MFs as those detected above (Bonatto, Bica \& Santos Jr. \cite{BBS2005}). Thus, the flat MFs 
and the presence of the MF break in clusters of any age and mass (Table~\ref{tab3} and Fig.~\ref{fig10}) 
must essentially reflect the effects of the internal dynamics of clusters on the MFs and/or some 
fundamental property of the IMF associated to different conditions in star formation. Kroupa 
(\cite{Kroupa2002}) points out that systematic differences in the IMF should appear between low-density 
environments and high-density regions. To investigate this issue we test relations involving parameters 
of the clusters which present the MF break. The results are in Fig.~\ref{fig11}.

Systematic MF flattening with time due to dynamical evolution is manifest in panel (a) of 
Fig.~\ref{fig11} which shows that the overall MF slope (in the mass range $\rm m\leq\mb$) 
decreases with age, particularly for the less-massive clusters. A similar trend is also present 
in the core MF slope (panel (d)). The decrease of core and overall MF slopes with age was present 
in the mass range $\rm m\geq\mb$\ (Fig.~\ref{fig7}, panels (j) and (d)).

Similarly to the mass range $\rm m\geq\mb$, core and overall MF slopes correlate as well (panel (g)), 
particularly in the less-massive clusters. The scatter hinders conclusions on dependences of MF slope 
on mass density in the core (panel (h)) and in the whole cluster (panel (e)). Except for the populous 
cluster NGC\,2477, \mb\ seems to decrease with overall cluster mass (panel (b)). The same trend occurs 
for core mass, since core and overall mass are tightly correlated (panel (j) in Fig.~\ref{fig9}). Within
the scatter we do not find correlations of \mb\ with mass density, both in the core (panel (f)) and in 
the whole cluster (panel (c)). However, \mb\ seems to decrease with cluster age for the clusters older 
than 1\,Gyr (panel (i)).

MF slopes in the mass range $\rm m\leq\mb$\ (break region) and $\rm m\geq\mb$\ correlate, both 
in the core (panel (j)) and particularly in the whole cluster (panel (k), correlation coefficient 
0.96). This suggests a common origin for the mechanisms responsible for the excess MF flattening in 
$\rm m\leq\mb$\ and that in $\rm m\geq\mb$.

\begin{figure} 
\resizebox{\hsize}{!}{\includegraphics{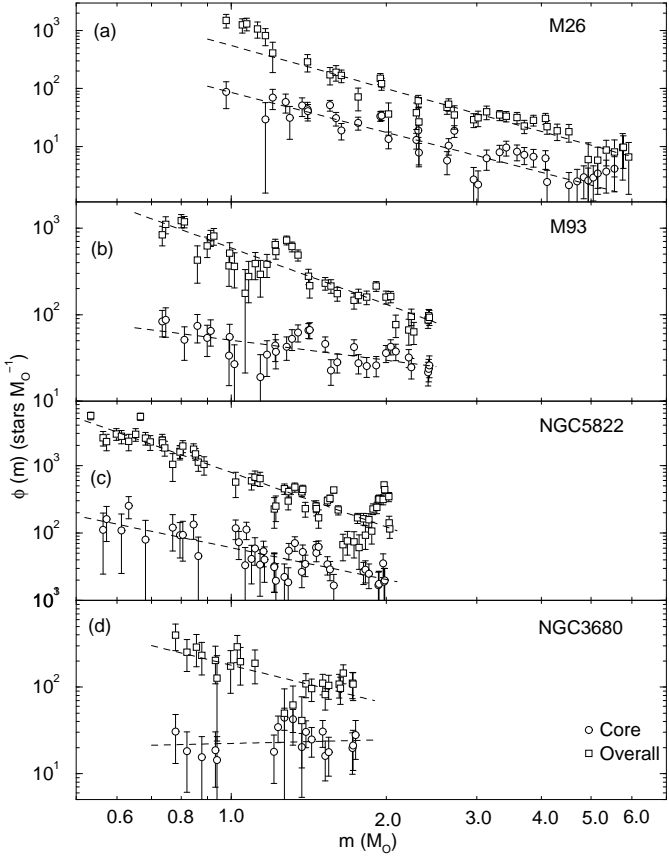}}
\caption[]{Open clusters with no clear evidence of a MF break. Because of the distance from the
Sun, the MF of M\,26 in panel (a) reaches down to $\rm m\approx0.98$\,\ms.}
\label{fig12}
\end{figure}

In Fig.~\ref{fig12} we show the MFs of the clusters which can be fitted by a single power-law. 
Because of the relatively large distance from the Sun (Table~\ref{tab1}), the low-mass end of the
observed MF of M\,26 (panel (a)) occurs at $\rm m\approx0.98$\,\ms, nearly the same value in which 
the MF break usually begins. Deeper observations are needed to access the MF break region in
M\,26. For M\,93 (panel (b)) and NGC\,3680 (panel (d)) the single power-law extends down to 
$\approx0.7$\,\ms\ and $\approx0.8$\,\ms\ respectively, with no evidence of a break. This is 
particularly true for the MF of NGC\,5822 (panel (c)), which extends to $\approx0.53$\,\ms\ 
fitted with a single power-law. The MF of NGC\,188 is not shown in Fig.~\ref{fig12} because the
distance from the Sun ($\ds\approx1.7$\,kpc) combined to the old age ($\approx7$\,Gyr) produce
a MF restricted to the narrow mass range $\rm 0.9\leq m(\ms)\leq1.1$. The mass range is too
narrow for any inferences on the presence of a MF break in this old open cluster. 

Recall that the MFs of the present cluster sample have been built 
by taking into account the 2MASS 99.9\% Completeness Limit (Sect.~\ref{MF}). Consequently, the 
low-mass end of each MF presented in this paper corresponds to the same apparent magnitudes, 
$\jj=15.8$, $\hh=15.1$\ and $\ks=14.3$. In this sense, if the MF break was artificially created 
by detection incompleteness in the faint magnitude limit, it should be present in the MFs of all 
clusters, irrespective of the corresponding low-mass limit. This is not the case of the MFs of 
M\,26, M\,93, NGC\,5822 and NGC\,3680 (Fig.~\ref{fig12}), in which the low-mass limit varies from 
$\approx0.5$\,\ms\ (NGC\,5822) to $\approx1.0$\,\ms\ (M\,26).

he presence of the MF break is not associated to cluster age and mass or concentration parameter,
at least in the studied clusters. However, the value of \mb\ seems to decrease with cluster mass.
We conclude that the MF break for $\rm m\leq1$\,\ms\ is definitively present in NGC\,2516, NGC\,2287, 
M\,48, NGC\,2477, IC\,4651 and M\,67. On the other hand it is absent, at least for the mass range 
$\rm m\geq0.7$\,\ms, in M\,93, NGC\,5822 and NGC\,3680. Further research accessing low-mass stars
in a statistically significant sample of open clusters is necessary to settle the issue whether 
or not the MF break is present in all cases, and thus test the hypothesis of the universal character 
of Kroupa's IMF. 

\subsection{Mass segregation and low-mass stars evaporation}

A comparison of the core and overall MF slopes in Table~\ref{tab3} may help us to evaluate the amount 
of mass segregation and low-mass stars evaporation which have already took place in the clusters. Except 
for M\,26, the core MF slope in the remaining clusters is much flatter than the overall one. In the 
young and rather massive (Table~\ref{tab3}) cluster M\,26 both MF slopes correspond essentially to the 
standard Salpeter one, within uncertainties. This suggests that mass segregation and low-mass stars 
evaporation have not yet had time to produce observable changes neither in the internal structure of 
M\,26 nor in the MFs. This fact is consistent with the overall relaxation time $\rm\tr\sim180$\,Myr, 
which is larger than the cluster age ($\sim70$\,Myr). On the other hand, the similarity between the core 
and overall MF slopes in M\,26 occurs despite the much shorter core relaxation time of $\rm\tr\sim3.2$\,Myr. 
This apparent contradiction reflects the complex way in which internal dynamical evolution relates to age 
and relaxation time, and in particular the different evolution time-scales associated to the core and 
overall regions.

In order to investigate the relationship between relaxation time and cluster age with dynamical evolution,
and to estimate the corresponding effects on MFs we calculate for each cluster the evolutionary parameter 
($\tau$), which is defined as the ratio of the cluster age to the relaxation time, $\rm\tau=age/\tr$. We 
take advantage of the 2MASS spatial coverage to examine separately the evolution of the core and overall 
regions. In Table~\ref{tab5} we list \tr\ and $\tau$ for each cluster. 

Considering separately massive and less-massive clusters, we found that core and overall relaxation 
times are linearly related. For the massive clusters we derive the relation
$\rm\tr(overall)\sim(89\pm15)\times\tr(core)$ (correlation coefficient of $\rm CC\approx0.94$) and
an average of $\sim120$. For the less-massive ones 
$\rm\tr(overall)\sim(56\pm20)\times\tr(core)$ ($\rm CC\approx0.85$) and an average of $\sim80$.  

\begin{table}
\caption[]{Relaxation time and evolutionary parameter.}
\label{tab5}
\renewcommand{\tabcolsep}{1.5mm}
\begin{tabular}{lccccc}
\hline\hline
&\multicolumn{2}{c}{Core}&&\multicolumn{2}{c}{Overall}\\
\cline{2-3}\cline{5-6}\\
Cluster&\tr&$\tau$&&\tr&$\tau$\\
&(Myr)&&&(Myr)\\
 ~~~(1)&(2)&(3)&&(4)&(5)\\
\hline
M\,26&$3.2\pm2.0$&$22\pm14$&&$178\pm116$&$0.4\pm0.3$\\
NGC\,2516&$0.4\pm0.1$&$365\pm74$&&$94\pm24$&$1.7\pm0.4$\\
NGC\,2287&$0.8\pm0.2$&$212\pm46$&&$40\pm10$&$4.0\pm1.0$\\
M\,48    &$0.7\pm0.2$&$517\pm166$&&$48\pm28$&$7.0\pm4.5$\\
M\,93&$0.9\pm0.2$&$457\pm112$&&$147\pm93$&$2.7\pm1.7$\\
NGC\,5822&$2.4\pm1.3$&$423\pm234$&&$262\pm170$&$3.8\pm2.5$\\
NGC\,2477&$6.2\pm2.5$&$181\pm75$&&$597\pm344$&$1.9\pm1.0$\\
NGC\,3680&$0.2\pm0.1$&$7100\pm2700$&&$29\pm17$&$55\pm32$\\
IC\,4651 &$0.6\pm0.1$&$2854\pm455$&&$33\pm7$&$54\pm13$\\
M\,67    &$1.2\pm0.2$&$2611\pm402$&&$80\pm15$&$40\pm8$\\   
NGC\,188&$14\pm9$&$504\pm321$&&$631\pm400$&$11\pm7$\\ 
\hline
\end{tabular}
\begin{list}{Table Notes.}
\item The relaxation time is calculated according to 
$\rm\tr\approx0.12\left(\frac{N}{\ln N}\right)\left(\frac{R}{1pc}\right)\left(\frac{\sigma_v}
{1\,\kms}\right)^{-1} (Myr)$, where N is the total number of stars, R is the radius of
the region considered and $\sigma_v$ is the velocity dispersion. We adopted $\rm\sigma_v\approx3\,\kms$ 
(Binney \& Merrifield \cite{Binney1998}). $\rm\tau=age/\tr$\ is the evolutionary parameter. 
\end{list}
\end{table}

We assume that flat overall MF slopes can be accounted for by low-mass stars evaporation resulting from 
large-scale mass segregation and external effects such as tidal stripping by the Galactic gravitational 
field. In this case, overall MF flattening is expected to be associated to $\rm\tau(overall)$. In fact, 
in the clusters M\,26, NGC\,2516, NGC\,2287, M\,48, M\,93, NGC\,5822 and NGC\,2477, which have overall 
MF slopes similar to that of Salpeter (thus, little low-mass stars evaporation), $\rm\tau(overall)\leq7$. 
Consequently, the age of these clusters results smaller than $\rm\sim7\times\tr(overall)$ which turns out 
to be too short a time for the evaporation to produce significant changes in the overall MFs. In this sense 
the flat core MFs of NGC\,2516, NGC\,2287, M\,48, M\,93, NGC\,5822 and NGC\,2477 can be accounted for by 
mass segregation, since for these clusters $\rm\tau(core)\geq200$. The nearly Salpeter MF in the core of 
M\,26 is probably due to the small $\rm\tau(core)\sim22$.

On the other hand, NGC\,3680, IC\,4651 and M\,67, which have flat overall MF slopes, have $\rm\tau(overall)\geq40$. 
The very flat core MFs in these clusters result from $\rm\tau(core)\geq2600$. Thus, the core and
overall evolutionary parameters are large enough for the large-scale mass segregation and low-mass stars 
evaporation to have produced significant MF flattening in these clusters.

Although with $\rm\tau(overall)\sim11$ and $\rm\tau(core)\sim504$, the oldest cluster in this work 
NGC\,188 doesn't seem to fit in the above scenario, because its overall MF slope is $\chi\sim1.9$. 
However, because of the old age and large distance from the Sun we could access the MF only over the 
reduced range $0.9\,\ms - 1.0\,\ms$, which produced large uncertainties in the MF slope. 

\begin{figure} 
\resizebox{\hsize}{!}{\includegraphics{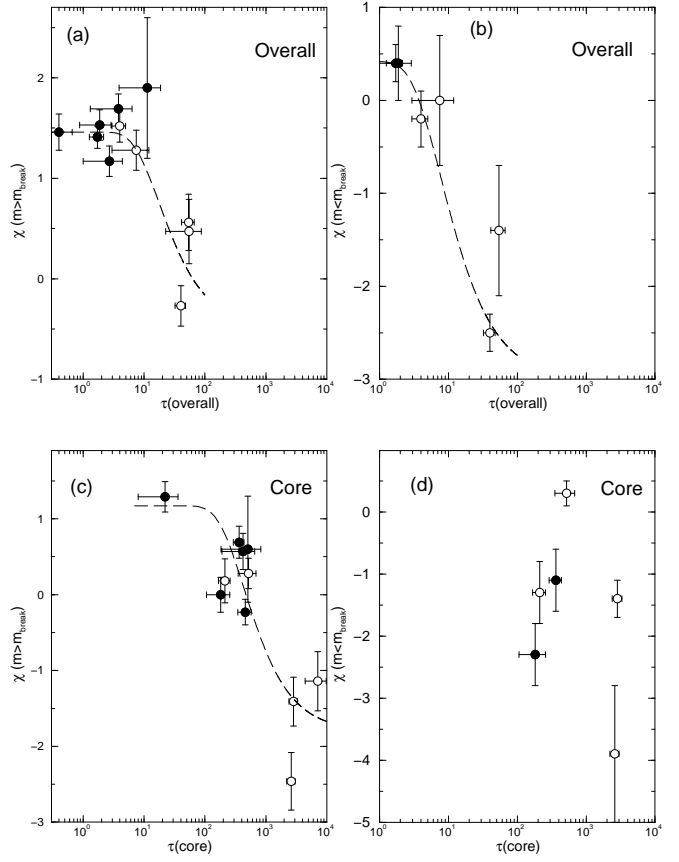}}
\caption[]{MF slopes as a function of the evolutionary parameter $\tau$. Left panels: MFs in the 
range $\rm m\geq\mb$; right panels: MFs in the range $\rm m\leq\mb$. Top panels: the scale in the 
abscissa is based on the overall $\tau$; bottom panels: abscissa in terms of core $\tau$. Dashed lines 
in panels (a), (b) and (c): tentative fit with $\rm\chi(\tau)=\chi_0-\chi_1\exp{\left(-\tau_0/\tau\right)}$. 
Symbols as in Fig.~\ref{fig7}.}
\label{fig13}
\end{figure}

We summarise the evolutionary scenario of core and overall MFs with respect to $\tau$\ in Fig.~\ref{fig13}. 
The overall MF slopes (for $\rm m\geq\mb$) of the massive clusters (panel (a)) do not change significantly 
for $\rm0.4\leq\tau(overall)\leq11$. On the other hand, the overall MFs of the less-massive clusters begin 
to flatten for $\rm\tau(overall)>7$, which means that evaporation effects tend to become appreciable on 
the MFs when the age of the cluster becomes larger than $\rm\sim7\times\tr(overall)$. This occurs probably 
because of the looser nature of the less-massive clusters. 

Because of high density and small dimensions (and thus small \tr), core MFs tend to flatten because of 
mass segregation for $\rm\tau(core)>22$ (M\,26) and $\rm\tau(core)\leq150$, in massive and less-massive 
clusters alike (panel (c)). To first order we take $\rm\tau(core)=100$ as representative of the value for
which core MF flattening becomes significant. 

Based on the discussion above we can express the time it takes for significant changes in the MFs (in the
range $\rm m\geq\mb$) to be produced as $\rm\Delta t=\tau\times\tr$. In the less-massive clusters the core 
and overall MFs begin to flatten at $\rm\tau(core)\sim100$ and $\rm\tau(overall)\sim7$, respectively. 
Taking into account the average ratio $\rm\tr(overall)/\tr(core)\approx80$, we find $\rm\Delta 
t(overall)\sim6\times\Delta t(core)$. We conclude that appreciable slope flattenings in the overall MFs 
of the less-massive clusters take on average $\sim6$ times longer to occur than in the core MFs. 

The value of $\rm\tau(overall)$ in which the overall MFs of massive clusters begin to flatten is not 
clear in panel (a) of Fig.~\ref{fig13}. Thus, assuming it as $\rm\tau(overall)\sim11\ (NGC\,188)$ for
simplicity, and considering the average ratio $\rm\tr(overall)/\tr(core)\approx120$, we find that
slope flattenings in the overall MFs of massive clusters take on average $\sim13$ times longer to occur 
than in the core MFs.

In the mass range $\rm m\leq\mb$\ the overall MFs (panel (b)) flatten for $\rm\tau(overall)\geq2$,
which indicates that in this mass range dynamical evolution effects are effective in flattening the
MFs quite rapidly. The scatter in panel (d) does not allow to draw conclusions on the behaviour 
of the core MFs. 

The discussion above (and the apparent dependence of $\chi$ on $\tau$ in Fig.~\ref{fig13}) suggests that 
the MF slopes undergo an exponential decay with $\tau$. Despite the reduced number of points we tentatively 
applied a fit with the empirical function $\rm\chi(\tau)=\chi_0-\chi_1\exp{\left(-\tau_0/\tau\right)}$ to 
the core and overall $\chi$ distributions (panels (a), (b) and (c) in Fig.~\ref{fig13}), combining massive 
and less-massive clusters. For $\rm\chi(overall)$ we derived $\chi_0=1.46\pm0.10$, $\chi_1=1.9\pm0.7$ and 
$\tau_0=18\pm14$, with a correlation coefficient $\rm CC=0.88$. For $\rm\chi(core)$ we found $\chi_0=1.17\pm0.23$, 
$\chi_1=3.0\pm0.7$ and $\tau_0=439\pm156$, with $\rm CC=0.88$. The resulting fits are shown in Fig.~\ref{fig13}. 
For the overall $\chi$ in the range $\rm m\leq\mb$ (panel (b) we derive $\chi_0=0.42\pm0.27$, $\chi_1=3.4\pm0.4$ 
and $\tau_0=8\pm4$, with $\rm CC=0.98$. These analytical relations can be used to compare core and overall
MF flattening time scales. According to the above relations the overall MFs of the less-massive clusters 
would flatten by $\Delta\chi=0.35$ in $\sim4\times$ the time it would take for the core MF to flatten by 
the same amount. In the massive clusters this ratio would be $\sim6$. Both values basically agree with our 
previous estimates (see above). 

In the same way, the overall MFs in the range $\rm m\leq\mb$ flatten by $\Delta\chi=0.35$ in $\sim1/3$ of 
the time the $\rm m\geq\mb$ overall MFs flatten by the same amount. This is consistent with the fact that 
on the way towards energy equipartition, stars in the range $\rm m\leq\mb$ should have, on average, a larger 
velocity dispersion ($\rm\sigma_v\sim1/\sqrt{m}$) than those in the range $\rm m\geq\mb$ and thus, they may 
leave out the cluster first. 
 
A larger number of clusters is necessary to first check the existence of this analytical representation of 
$\chi(\tau)$ and then to derive more accurate values of $\rm\chi_0$, $\rm\chi_1$ and $\rm\tau_0$, and finally, 
discuss further implications.

\section{Evidence of an open cluster Fundamental Plane}
\label{FP}

In Sect.~\ref{MPA} we discussed relations involving pairs of intrinsic open cluster parameters. We 
found a few correlations among core and overall parameters, while in some plots massive and less-massive 
clusters are clearly segregated. This raises the question whether one of these pairs might be related 
to a third parameter, which could reflect the existence of a fundamental plane (FP) of open cluster 
parameters, similar to that of the elliptical galaxies (Dressler et al. \cite{Dress1987}; Djorgovski 
\& Davis \cite{DD1987}). Elliptical galaxies populate a plane in the 3D-parameter space defined by
the luminosity (L), effective surface brightness ($\rm\mu_e$) and velocity dispersion ($\rm\sigma_v$).
The plane comes about because as a consequence of the virial theorem a well-defined relationship
between mass (M), $\rm\mu_e$\ and $\rm\sigma_v$\ is expected to occur in elliptical galaxies. Scatter 
in the plane depends essentially on the function relating the ratio M/L with L and $\rm\mu_e$\
(Lucey, Bower \& Ellis \cite{Lucey1991}). There is suggestive evidence that the planes of the
ellipticals in the Coma cluster and that in the Virgo are not parallel (Lucey, Bower \& Ellis 
\cite{Lucey1991}).

Although complete virialization in open clusters is an idealization, as clusters of different masses 
age they may reach advanced dynamical states. Indeed, kinetic theory and numerical simulations predict 
that N-body encounters in multi-mass models tend to produce energy equipartition within a time-scale 
of the order of the relaxation time (de la Fuente Marcos \& de la Fuente Marcos \cite{delaF2002}). 
Except in M\,26, the overall relaxation times for the remaining clusters in the present work are
significantly smaller than the cluster age. Core relaxation times are indeed much smaller than
the ages (Table~\ref{tab5}). In this sense, varying levels of relationship between open cluster 
parameters are expected to occur, particularly in the old clusters. 

Open clusters do differ of elliptical galaxies particularly in terms of stellar population (absolute 
numbers and age), linear dimension and dynamical evolutionary state. Thus, the existence of an
equivalent FP in open clusters raises some questions. {\em (i)} Is the open cluster FP related to 
dynamical evolution? {\em (ii)} Does the massive/less-massive cluster segregation (Sect.~\ref{MPA}) 
still hold in the FP? {\em (iii)} Does Galactocentric distance affect the FP?
A more comprehensive sample of open clusters is essential to answer these questions. 

\begin{figure} 
\resizebox{\hsize}{!}{\includegraphics{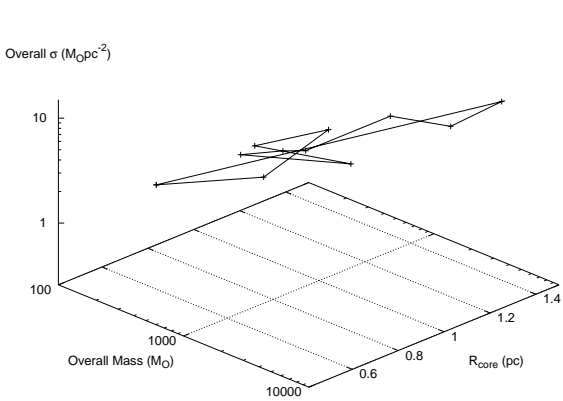}}
\caption[]{Evidence of a fundamental plane of open cluster parameters. The distribution of overall 
mass (x-axis), core radius (y-axis) and projected overall mass density (z-axis) essentially defines 
a plane in this 3D projection.}
\label{fig14}
\end{figure}

In the case of the present open clusters (Tables~\ref{tab2} and \ref{tab3}) we found that the overall 
mass and core radius are effectively related to the projected overall mass density. In Fig.~\ref{fig14} 
we show a 3D-projection of those three parameters, where one can see that they distribute nearly in a 
plane. A similar relation occurs for overall mass, core radius and projected core mass density.

At the present stage the number of open cluster parameters is of the same order as the 
number of objects already analyzed by the techniques presented in the previous sections. We 
are using the same techniques to derive parameters of a larger number of clusters with 
significantly different properties, such as age, mass, core and limiting radii, densities, 
Galactocentric distance, etc. Our goal is to apply a principal component analysis on this
comprehensive cluster sample not only to search for correlations among parameters, but to 
better explore the possibility of an open cluster FP as well. 

\section{Concluding remarks}
\label{Conclu}

We analysed 11 nearby open clusters with ages in the range 70\,Myr to 7\,Gyr and masses in the range 
$\approx400$\,\ms\ to $\approx5\,300$\,\ms\ based on \jj, \hh\ and \ks\ 2MASS photometry. The 
clusters are M\,26, NGC\,2516, NGC\,2287, M\,48, M\,93, NGC\,5822, NGC\,2477, NGC\,3680, IC\,4651, 
M\,67 and NGC\,188. Radial density profiles and mass functions were built after taking into account 
the field contamination. As a consequence we derived a homogeneous set of parameters associated to 
the structure and stellar and dynamical evolution of the clusters. 

The method is illustrated in detail by analysing the populous open clusters NGC\,2477 and NGC\,2516 
for the first time in the near-infrared. For NGC\,2477 we derive an age of $1.1\pm0.1$\,Gyr, distance 
from the Sun $\ds=1.2\pm0.1$\,kpc, core radius $\rc=1.4\pm0.1$\,pc, limiting radius $\rl=11.6\pm0.7$\,pc 
and total mass (extrapolating the MF down to the H-burning mass limit, 0.08\,\ms) 
$\mtot\approx(5.3\pm1.6)\times10^3$\,\ms. The MF slope varies significantly in this cluster, being 
flat ($\chi\approx0.0$) in the core ($\rm 0.0\leq r(pc)\leq1.4$) and steep ($\chi\approx3.3$) in the 
outskirts ($\rm 5.2\leq r(pc)\leq11.7$). The overall MF has a slope $\chi\approx1.5$. The number-density 
of evolved stars (with respect to MS stars with mass down to $0.08\,\ms$) in the core is about 4 
times as large as in the cluster as a whole. In the halo the number-density of low-mass stars is
$\sim4$\ times as large as that of the evolved stars. These facts reflect the effects of large-scale 
mass segregation in NGC\,2477. For NGC\,2516 we derive an age of $160\pm10$\,Myr, $\ds=0.44\pm0.02$\,kpc, 
$\rc=0.6\pm0.1$\,pc, $\rl=6.2\pm0.2$\,pc and $\mtot\approx(1.3\pm0.2)\times10^3$\,\ms. Similarly to 
NGC\,2477, mass-segregation effects in NGC\,2516 are reflected in the spatial variation of the MF 
slopes, $\chi\approx0.7$ in the core ($\rm 0.0\leq r(pc)\leq0.6$) and $\chi\approx1.8$ in the
outskirts ($\rm 1.6\leq r(pc)\leq6.2$). The overall MF has a slope $\chi\approx1.4$. The overall
MFs of NGC\,2477 and NGC\,2516 have slopes similar to a standard Salpeter IMF ($\chi\approx1.35$).

Six of the 11 clusters present a break around 1\,\ms\ followed by a sharp flattening in 
the MF in the mass range $\approx0.5$\,\ms\ to $\approx1$\,\ms. The mass range where we detected
the MF break in open clusters basically coincides with previous results based on star counts in 
the solar neighbourhood (Kroupa \cite{Kroupa2001}, \cite{Kroupa2002}). In the clusters analysed 
here the MF break is not associated to mass or concentration parameter. The presence of the MF
break in all clusters would be consistent with the universal character of the IMF (Kroupa 
\cite{Kroupa2002}). However, 3 clusters in our sample do not present the MF break, at least for the mass 
range $\rm m\geq0.7\,\ms$. This does not exclude the possibility that in some clusters the MF 
break occurs at lower masses or that dynamical effects may somehow damp out features in the MFs. 
To settle these issues we are analyzing a more statistically significant sample of open clusters. 

We investigated the dependence of MF slopes on the evolutionary parameter $\rm\tau=age/\tr$. We
found that dynamical effects begin to produce substantial flattening in the core MFs of massive and
less-massive clusters  when the cluster age is $\rm\sim100\times\tr(core)$. In the overall MFs the 
changes become noticeable when $\rm\sim7\times\tr(overall)$ and $\rm\sim11\times\tr(overall)$,
respectively for the less-massive and massive clusters. Considering the linear relationship 
between core and overall relaxation times, we conclude that appreciable slope flattening in the 
overall MFs of the less-massive clusters take $\sim6$ times longer to occur than in the core MFs. 
In the massive clusters they take a time $\sim13$ times longer. We found that MF slopes vary with 
$\tau$ according to the empirical relation $\rm\chi(\tau)=\chi_0-\chi_1\exp{\left(-\tau_0/\tau\right)}$.

We also searched for relations of cluster parameters with cluster age and Galactocentric distance.
The main results are: {\em (i)} cluster size correlates both with cluster age and Galactocentric 
distance; {\em (ii)} because of size and mass scaling, core and limiting radii, and core and overall 
mass correlate as well; {\em (iii)} core radius and overall mass are correlated, but massive 
($\rm m\geq1\,000\,\ms$) and less-massive clusters follow separate, nearly parallel paths on the 
plane; {\em (iv)} the MF slopes of massive clusters are restricted to a narrow range, while those
of the less-massive clusters distribute over a wider range; {\em (v)} core and overall MF slopes
are correlated.

We found relations involving three parameters simultaneously which may suggest a fundamental plane
of open cluster parameters. The relations involve overall mass, core radius, and projected overall 
mass density or projected core mass density, which are those more closely related to the parameters
involved in the Fundamental Plane of ellipticals.

As prospective work we are carrying on the present analysis for a larger sample in order to draw 
more quantitative conclusions on the fraction of open clusters which present a break in the MF, 
the reason why some do not have the break, and on correlations of \mb\ with other cluster 
parameters, further exploring as well the possibility of a fundamental plane. 

\begin{acknowledgements}
We thank the anonymous referee for helpful suggestions.
This publication makes use of data products from the Two Micron All Sky Survey, which 
is a joint project of the University of Massachusetts and the Infrared Processing and 
Analysis Center/California Institute of Technology, funded by the National Aeronautics 
and Space Administration and the National Science Foundation. We also made use of the 
WEBDA open cluster database. We acknowledge support from the Brazilian Institution CNPq.
\end{acknowledgements}

%

\end{document}